\documentclass{ws-rv9x6}
\usepackage{subfigure}   % required only when side-by-side / subfigures are used
\usepackage{ws-rv-thm}   % comment this line when `amsthm / theorem / ntheorem` package is used
\usepackage{ws-rv-van}   % numbered citation & references (default)
\usepackage{subfig}
\usepackage{float}
\usepackage{slashed}
\usepackage{rotating}
\usepackage{multirow}
\usepackage{psfrag}
\usepackage{tabularx}
\makeindex
\newcommand{\newc}{\newcommand}
\newc{\beq}{\begin{equation}}
\newc{\eeq}{\end{equation}}
\newc{\barr}{\begin{eqnarray}}
\newc{\earr}{\end{eqnarray}}
\newc{\bea}{\begin{eqnarray}}
\newc{\eea}{\end{eqnarray}}

\def\lambf{\mbox{\boldmath $\lambda$}}

\def\xibf{\mbox{\boldmath $\xi$}}

\def\sbf{\mbox{\boldmath $\sigma$}}

\def\etabf{\mbox{\boldmath $\eta$}}

\begin{document}

%\tableofcontents
%\listoftables
%	\setcounter{chapter}{0}
\chapter[Symmetries in subatomic six-quark systems]{Symmetries in subatomic six-quark fermions}\label{ra_ch1}
\author[J. D. Vergados and D. Strottman]{J.D. Vergados\footnote{Physics Department, University of Ioannina, Ioannina, Gr 451 10, Greece.} and D. Strottman\footnote{Theoretical Division, LANL, Los Alamos, NM 87544, USA,  and FIAS, Goethe University, Frankfurt, Germany} }
%\aindx{Author, F.}                       % author index entry
%\address{$^1$LANL, Los Alamos, NM 87544, USA, $^2$Physics Department, University of Ioannina, Ioannina, Gr 451 10, Greece}
%\address{$^2$ University of Ioannina, Ioannina, Gr 451 10, Gree
\begin{abstract} 
The evidence of the existence of tetra-quark meson and   penta-quark fermion systems has revived interest in the hunt of other multi-quark systems. We explore the possibility  of six-quark cluster configurations with the two nucleon spin-isospin  quantum numbers existing  in the nuclear medium. The investigation of such clusters is feasible if one exploits all the symmetries involved, in particular 
orbital, color-spin  $SU(6)_{cs}$ and isospin SU(2). 
We investigate whether such clusters contribute  to neutrinoless double-beta decay mediated by heavy neutrinos or other exotic particles, since in conventional nuclear physics the relevant nuclear matrix elements are suppressed due to the presence of the nuclear hard core.
\end{abstract}
\body
%\section{Using Other Packages}
%The class file has...
%\begin{appendix}[Optional Appendix Title]
%\section{Sample Appendix}
%Text...
%\end{appendix}
%\begin{thebibliography}{9}        % for non BIBTeX users
%\bibitem{ams04} \AmS, \emph{\AmS-\LaTeX{} ...
%\end{thebibliography}
%\printindex[aindx]                % to print author index
%\printindex

\section{Introduction}
\label{sec:Intro}
%\bibliographystyle{ws-ijmpe}
%\bibliography{Bib}
%%%%%%%%%%%%%%%%%%%%%%%%%%%%%%%%%%%%%%%%%%%%%%%%%%%%%%%%%%%%%%%%%%%%%%%%%%%%%%%%%%%%%%
We first met Akito when he visited Rutgers and Stony Brook Universities in 1967-68 when we were both at Stony Brook.  In addition he visited Oxford  for several weeks where one of us (DS) was a Research Fellow as well as numerous intervening visits at several locations around the world.  His ideas and guidance then have laid the foundation for much work through the decades including parts of this paper.  We dedicate this paper to his memory.

Multi-quark systems are fermions containing more than three-quarks or mesons containing more pairs than one quark or one anti-quark. The first calculations\cite{Str78a,Str78b} of such systems employed the Jaffe version \cite{Jaf77,Jaf77a}  of the MIT bag model. Hints of such multi-quarks have since been seen observed experimentally, the penta-quark \cite{pentaq05,pentaq15,CernCP16,pentaq16a} 
%{\bf check} 
and the tetra-quark \cite{fourq10,fourq14}. For the latter see also the recent reviews \cite{pentaq15,CernCP16,fourq14,VerMou21}.
Since penta-quark and tetra-quark systems have already been found to exist, see \cite{pentaq15,CernCP16, Xiao, fourq10, Ikeno} and references therein, one wonders are there not multi-quark configurations in the nucleus?  And what signatures might reveal their presence? In this paper we will examine the possibility of the presence of six-quark clusters in the nucleus, a  more complex problem. Such clusters, if present with a reasonable probability in the nucleus, may contribute to various processes such as neutrinoless double-beta decay mediated by heavy neutrinos or other exotic particles. In nuclear physics the relevant nuclear matrix elements are suppressed due to the presence of the nucleon-nucleon hard core. In the presence of such clusters, however, the interacting quarks are in the same hadron and one can have a contribution even in the case of a $\delta$-function interaction \cite{JDV85}. 

The paper is organized as follows: In section  \ref{sec:6qmodel} we present the mathematical  model  describing   six-quark states, in \ref{sec:FynCal} we exhibit the techniques to obtain the wave functions, in section \ref{sec:MulttqME} we discuss the structure of the $q^6$ eigenstates, in \ref{sec:BaryonClusters} we expand the six-quark states into two  clusters of the form $q^3 \times q^3$, in \ref{sec:MicBSM} we examine the mixing of the six-quark states with  two-particle shell  model states, and in \ref{sec:Ccfpbeta} we consider the role of six-quark admixtures  in the  $0\nu\beta\beta$  decay, mediated by heavy Majorana neutrinos. Our conclusions are presented in section \ref{sec:ConRem}.

\section{Model to Describe  a Six-Quark State}
\label{sec:6qmodel}

Symmetries play a crucial role in reliably estimating the probability of calculating such six-quark clusters in the nucleus.  Already in 1937 Wigner \cite{Wigner,Wigner1} introduced his supermultiplet model which uses the $SU(2)_\sigma \times SU(2)_\tau \subset SU(4)$.   Wave functions described by such a labelling can naturally describe selection rules for operators that describe beta decay and most electromagnetic transitions.  

Akito Arima and colleagues\cite{Inoue1, Akiyama1, Akiyama2} developed a shell model code to describe nuclei A=18-22 based on the 
SU(3) model. We shall use a similar approach albeit with a generalized group structure to describe the system of six-quarks.    In a previous article\cite{Str-Ver} we developed the formalism needed in the evaluation of the energy of  
six-quark  cluster configurations that can arise in a harmonic oscillator basis of up to $2\hbar\omega$ excitations. The 
symmetries that were found useful for this purpose are those from the Jaffe Model:   the combined colour-spin symmetry 
$SU_c(3)\times SU_\sigma (2) \subset SU_{cs} (6)$,  and the isospin symmetry $SU_I (2)$.  
However, unlike the Jaffe model and the $q^4\bar{q}$ calculations  in which the radial wave functions were a spherical 
Bessel function, here we employ three-dimensional harmonic oscillator wave functions to confine the quarks.  This allows us  to introduce an additional symmetry, $SU_r (3)$, for the spatial degrees of freedom.  In this sense our calculations are not  unlike those using a harmonic oscillator functions to describe a three-quark system; see ref. \cite{Richard} and references therein.

Building on the work of Jahn and collaborators,\cite{JW,EHJ,Elliott-Flowers} in 1953  Elliott\cite{Elliott-1958, Elliott-Harvey,Harvey-Adv}   introduced the $SU(3$) group to characterize states in light nuclei and showed how characteristics of rotational bands could develop.   Akito Arima and colleagues\cite{Inoue1, Akiyama1} advanced this work by implementing the technology in a computer code.  This effort was extended to other nuclei in the 1s0d shell \cite{Arima-DDS1,Arima-DDS2,DDS-Mg} by Manakos and the Darmstadt group as well asto the 1p0f shell\cite{DDS-Oxford}.   
Necessary SU(3) recoupling coefficients and Wigner coefficients in algebraic form were calculated by Hecht\cite{Hecht-1} and Vergados\cite{Vergados}.  Subsequently, computer codes for the recoupling coefficients and Wigner coefficients were published by Akiyama and Draayer\cite{Draay-A,Draay-B}.  Nine-$(\lambda \mu)$ coefficients were defined by Millener \cite{John-recoupling}.  The $SU(3)\times SU(2) \subset SU(6)$ coupling coefficients were calculated were published in ref \cite{dds1,dds2}.  These results are critical for the calculations described in this paper.

\subsection{Basis States} 

 In the following $[f] = [f_1 f_2 \cdots f_6]$ will denote the Young tableaux that labels an irreducible representation of $SU(6), SU(3),$ and and $S_n$\cite{Weyl1}.   Following Elliott if $[g_1g_2g_3]$ denotes a Young tableaux labelling a representation of $SU(3)$, then the quantities
\begin{eqnarray}
\lambda &=& g_1 - g_2 \nonumber \\
\mu &=& g_2 - g_3
\end{eqnarray}
provide a convenient labelling scheme useful for many-quark states.  The $SU(3)$ representations, be they flavour, colour, or spatial, are labelled by $(\lambda \mu)$.   Our notation is briefly described in references \cite{Str-Ver,dds1}.

In this notation the dimension of an $SU(3)$ representation $(\lambda \mu)$ is
$$d_{(\lambda\mu)} = \frac{1}{2} (\lambda + 1)(\mu +1) (\lambda+\mu+2).$$
A colour singlet is $(0 0)_c$, a colour octet $(1 1)_c$; a single quark carries the $SU(3)_c$ representation $(1 0)_c$.  Removing a single quark from a colour singlet $q^6$ state leaves a $q^5 \;(0 1)_c$ state.

As described in our previous paper\cite{Str-Ver} all basis states are products of four components: space, colour, spin, and  isospin; the colour and spin will be combined in colour-spin  wave function described by $SU(6)_{cs}$. This a generalization of LS-coupling shell model codes that described 0p, 1s0d, and 1p0f shell nuclei such as those of Arima {\it et al}.   We assume  the six-quarks must be a color singlet.     The isospin will be described by $SU(2)_I$ and the spatial wave functions -- being three-dimensional harmonic oscillator functions -- will be labelled by $S(N), [f]_r$, and $SU(3)_r, (\lambda \mu)_r$.  The $SU(6)_{cs}$ representations relevant to this work and
their spin and isospin content are shown in Table~\ref{tab:6qIR}. Also, the $\tilde{f}$ indicates the representation $[\tilde{f}]$ is contragradient 
to that of $[f]$.  The $SU(N)$ coefficients are usually referred to as fractional-parentage coefficients 
following the work of Racah\cite{Racah2,Talmi}.

\subsection{$0\hbar \omega$ Excitations}

 In the lowest energy basis state all six-quarks are assumed
to be in the $0s$ state.  The wave function is then
\begin{equation}
| (0s)^6 [6]_r\; [f]_{cs}(0 0)_c \;S \;I=1>\; .\label{eq:basis0s6}
\end{equation}
The orbital angular momentum is zero. Since the spatial wave function is totally symmetric, the product of
the colour-spin and isospin wave functions must be antisymmetric.  For
$I=1$ the isospin representation is necessarily$[42]_I$; from Table~\ref{tab:6qIR}
one sees the spin can be either 0 or 2.  

The former is diproton-like
but in the nucleus, one cannot exclude the S=2 possibility.
Note that the six-quarks can also have isospin two or three as well as
one or two.  One may think of these states as having $\Delta$
admixtures.  However,  if one decomposes the $I=0$ six-quark wave
function into two hadrons as we shall do in section \ref{sec:BaryonClusters} there will be $\Delta$ admixtures.

\subsection{One Particle, $2\hbar\omega$ Excitations}

These positive parity states are formed by promoting a $0s$ quark to the $1s0d$ shell.  The wave function  is 

\barr
|&&
| (0s)^5 [5]_r\; [f]_{cs}(0 1)_c\; L_1=0\;\;S_1 \;I_1 \;\times\nonumber\\
&&1s/0d\; [1]_{cs}(1 0)_c\;L_2\; S_2\;I_2=\frac{1}{2}\; \frac{1}{2};\; (0
0)_c\; L=L_2\;S \;J\;I=1>.\nonumber\\
\label{eq:basis0s5}
\earr

Depending on the total isospin of the six quarks, the five-quark wave function  can have  $I_1=\frac{1}{2}$ or 
$I_1=\frac{3}{2}$.  For the five-quark state with $I_1=\frac{1}{2}$,  $S_1$ can have the values $ \frac{1}{2},\frac{3}{2},\frac{5}{2}.$  If the isospin, $I_1$, is $\frac{3}{2}$, the $SU(2)$ representation is
$[41]_I$; the conjugate $SU(6)_{cs}$ is then $[2111]_{cs}$ which allows $S_1 = \frac{1}{2},\frac{3}{2}.$ The angular momentum, $L_2$, of the $(1s0d)$ quark can be either 0 or 2.
These states can combine to form a variety of total $L, S, J, I$.

\begin{table}\centering
	
	\begin{tabular}{|c|rr|c|}
		\hline $[f_{cs}]$ & $(\lambda \mu )_c $&$\;S$& I\\\hline
		\hline
		$[222]_{cs}$& $(0 0)_c$ &1, 3&0\\
		$[2211]_{cs}$& $(0 0)_c$ & 0, 2&1\\
		$[21^4]_{cs}$& $(0 0)_c$ & 1&2\\
		$[1^6]_{cs}$& $(0 0)_c$ & 0&3\\
		\hline
		\hline &&&\\
		$[221]_{cs}$ & $(0 1)_c $&$ \;\frac{1}{2}, \frac{3}{2},
		\frac{5}{2}$&$\frac{1}{2}$\\&&&\\
		\hline &&&\\$[2111]_{cs}$ & $(0 1)_c $&$\;\frac{1}{2},
		\frac{3}{2}$&$\frac{3}{2}$\\&&&\\
		\hline &&&\\$[1^5]_{cs}$ & $(0 1)_c $&$\;\frac{1}{2}$&$\frac{5}{2}$\\&&&\\\hline
		\hline $[22]_{cs}$ & $(0 2)_c$& 0 & 0\\
		& $(0 2)_c$& 2& 0\\
		& $(1 0)_c$ &1& 0\\\hline
		$[211]_{cs}$ & $(0 2)_c$& 1& 1\\
		& $(1 0)_c$ &0& 1\\
		& $(1 0)_c$ &1& 1\\
		& ($(1 0)_c$ &2& 1\\\hline
		$[1^4]_{cs}$ & $(0 2)_c$& 0& 2\\
		& $(1 0)_c$ &1& 2\\
		\hline\end{tabular}
	\caption[Relevant six-, five- and four-quark $SU(6)_{cs}$ representations]{ $SU(6)_{cs}$ representations for six, five, and four quarks that contain $(\lambda \mu
		)_c$ representations relevant to the calculation taken from Ref. \cite{dds1}.  Those
		$SU(6)_{cs}$ representations that have more than two columns are not
		allowed because the contragradient representation must belong to
		$SU(2)_I$.
	}
	\label{tab:6qIR}
\end{table}

\subsection{Two-Particle, $2\hbar\omega$ Excitations}

These states are formed by promoting two $0s$ quarks to the $0p$ shell.  The wave function can be written as

\barr
|&& (0s)^4 [4]_r\; [f]_{cs}(\lambda_c \mu_c)_c\; L_1=\;\;S_1 \;I_1
\;\times\nonumber\\
&& (0p)_r^2\;[f_r] (\lambda_r \mu_r)\;
\;[f_2]_{cs}\;(\mu_c\lambda_c)_c\; S_2\; \;I_2;\; (0 0)_c\; L\;S
\;J\;I=1>.\label{eq:basis0s4}
\earr

 The spatial symmetry the $0p^2$ wave functions, 
 $[f_r]$, may be either [2] or [11].  We must have
\begin{equation}
|space \times SU(6)_{cs} \times isospin >
\end{equation}
totally antisymmetric.   Note that with spatial symmetry
	included  which can be either symmetric or anti-symmetric,
	$[f]_{cs}$ no longer determines $]f]_{f}$ uniquely.

For the $0s^4$ states the spatial part is symmetric so we need deal only with
colour-spin and isospin.   The possible SU(2) isospin
representations are $[22] I=0$, $[31] I=1$, and $[4] I=2$.  The
corresponding $SU(6)_{cs}$ representations are $[22]_{cs},
[211]_{cs}$, and $[1111]_{cs}$, respectively.  The colour-spin content
is shown in Table~\ref{tab:6qIR}. 

By combining all these considerations, one produces the complete basis states for our problem.  It is important that one includes all possible basis states for reasons of projecting out spurious states; this will be explained below.  For $J=0, I=1$ there are 31 basis states of which 10 are $L=0,$ 11 are $L=1$, and 10 are $L=2$.

\section{Calculation of the Wave Functions }
\label{sec:FynCal}

The methods of calculation and relevant formulae were presented in ref.\cite{Str-Ver}.  Our  approach generalizes the earlier work of Jahn\cite{JW}, Elliott,\cite{EHJ,Elliott-Flowers},  Arima\cite{Inoue1}, and others.

%\section{$\beta\beta$ Process}
\subsection{Matrix Elements of the Interaction}
\label{sec:MEint}

In this work we shall use two different quark-quark interactions motivated by QCD and previous work.
The leading order of the one-gluon exchange between two quarks is $$\frac{\alpha_s}{b}\frac{1}{r_{12}}\lambf_1 \cdot \lambf_2.$$

The matrix elements of the SU(3) generators are
$$\langle \,(\lambda,\mu)|\;\lambf_1.\lambf_2|\,(\lambda,\mu)\,\rangle_c=\left \{\begin{array}{rr}\frac{2}{3},&(\lambda,\mu)_c=(2,0)_c\\-\frac{4}{3},&(\lambda,\mu)_c=(0,1)_c\\ \end{array} \right .  .$$
This factor must then be multiplied by the matrix elements of the radial components which may have an additional spin dependence.

The two-body matrix elements in the case of the one-gluon exchange  potential are given in terms units of $\alpha_s/b$.
Hadron studies seem to suggest a value of the strong coupling constant $\alpha_s$ to be a factor of five times larger than that employed in high energy physics\cite{VerMou21}. We adopt a value $\alpha_s=1$.

We also use an interaction employed by Ohta {\it et al}\cite{Ohta}:
\beq
V^A_{ij} = \frac{1}{4} \Big[ A e^{-\frac{r_{ij}^2}{\alpha^2}} + B r_{ij}^2 +C + K (1 + \frac{2}{3} (\sigma_1\cdot \sigma_2) ) \delta(r_{ij}) \Big]\;(\lambda_i\cdot \lambda_j ) .
\eeq
We use their set A of the parameters: A=3810, B=-12.5, C=-479.8, and K=-911.1.  

This interaction has the advantage of having an explicit spin-dependence.  The QCD interaction $\frac{1}{r_{12}}\lambda_1 \cdot \lambda_2 $ is spin-independent and because all our states are colour singlets and the radial matrix elements of $0s^2$ are identical, all $0s^6$ states are degenerate in energy.

\subsection{The Matrix Element  of a Tensor  Interaction.}

The one-gluon exchange potential also contains a tensor component.  The diagonal contribution of the tensor component is easily included as we explain. Again the matrix element  is diagonal in spin with $S=S'=1 $ So we find:
\barr
\langle[2] |V_{12}|[2]\rangle&=&\frac{4}{3} \sqrt{5}\mbox{ for }(\lambda,\mu)=(2,0),s=1,\nonumber\\\langle[1^2] |V_{12}|[1^2]\rangle&=&-\frac{8}{3} \sqrt{5}\mbox{ for }(\lambda,\mu)=(0,1),s=1\;.
\earr
We should mention that in this case the reduced spin matrix element is included, see ref\cite{HechtBook} for the definitions. The whole matrix element except for the radial integral as indicated below,  must be multiplied with the reduced matrix element of the orbital part times the Racah function $U(\ell 2js;\ell's)$;  that is a total factor of
\beq
f_{so}=\sqrt{2}\;U(\ell 2jS;\ell'S)\;\langle \ell' 0,20|\ell 0\rangle
\eeq
where the last factor is a Clebsch-Gordan coefficient,  $\ell, \ell'$ are the orbital angular momenta of the initial and final two quark states, $S=S'=1$ is their spin and $j$ their total angular momentum.

The radial integrals needed in this work must be multiplied by the following factors as shown in table 	\ref{tab:tensor}.\\
\begin{table}[H]
	\caption{The radial part $ <n\,\ell' | V | ss\,n \ell >$ of the tensor interaction  for  matrix $< i|V| j >$ . The matrix is symmetric. 
		\label{tab:tensor}}	
	$$
	\begin{array}{|c|c|c|cc}
	\hline
	i/j&4&5\\ \hline
	1&\frac{1}{\sqrt{2}} 0s(r) \, \,V(\sqrt{2}r)\, \, 0d(r)&-\frac{1}{\sqrt{2}} 0s(r) \,V(\sqrt{2}r)\, 0d(r)\\
	2&\frac{1}{\sqrt{2}} 1s(r) \,V(\sqrt{2}r)\, 0d(r)&-\frac{1}{\sqrt{2}} 1s(r) \,V(\sqrt{2}r)\, 0d(r)\\
	3&-\frac{1}{\sqrt{2}} 1s(r) \,V(\sqrt{2}r)\, 0d(r)&\frac{1}{\sqrt{2}} 1s(r) \,V(\sqrt{2}r)\, 0d(r)\\
	\hline
	\end{array}
	$$
	$$
	\begin{array}{|c|c|c|cc}
	\hline
	i/j&4&5\\ \hline
	4&\frac{1}{2} 0d(r) \,V(\sqrt{2}r)\, 0d(r)&-\frac{1}{2} 0d(r) \,V(\sqrt{2}r)\, 0d(r)\;\\
	5&-\frac{1}{2} 0d(r) \,V(\sqrt{2}r)\, 0d(r)\;&\frac{1}{2} 0d(r) \,V(\sqrt{2}r)\, 0d(r)\\
	\hline
	\end{array}
	$$
	$$
	\begin{array}{|c|c|c|cc}
	\hline
	i/j&6&8\\ \hline
	6& 0p(r) \,V(\sqrt{2}r)\, 0p(r)\;&0\\
	8&\;0&\;-\frac{1}{2} 0p(r) \,V(\sqrt{2}r)\, 0p(r)\;\\
	\hline
	\end{array}
	$$
\end{table}

One of the  motivations of introducing a tensor interaction was to explain the L = 2 component in the deuteron wave function.  It is also essential to explain electroweak transitions in nuclei.  In our case all the lowest $q^6$ states with isospin one has $L = S= 0$ and thus the tensor interaction does not contribute.  Its main contribution will - as in the case of the deuteron - be the admixing of the $L = 2$ states with the $L = 0$ states.  These admixtures will be left for subsequent work and will not be discussed in this paper.

The above expressions must be multiplied with the matrix element of the radial part one-gluon of the operator, which in the present  calculation will be taken $V(|{\bf r_1-r_2}|)=\frac{ \alpha_s}{|{\bf r_1-r_2}|}$ corresponding  to leading order to one-gluon exchange potential.

%%%%%%%%%%%%%%%%%%%%%%%%%%%%%%%%%%%%%%%%%%%%%%%%%%%%%%%%%%%%%%%%%%%\subsection{Determination of $\hbar \omega $ of the  Harmonic Oscillation Potential at the Quark Level}

We will consider two models:

{\it Model A}: A naive approach would be to use a modified version of the expression used in the nuclear shell model:
\beq
b_A=\frac{\hbar c}{m_N c^2 \hbar \omega_A}\Rightarrow \hbar \omega_q=\frac{(\hbar c)^2}{m_q c^2 b_N^2}=\frac{(\hbar c)^2}{(m_N /3)c^2 b_N^2},
\label{Eq:Homegacon}
\eeq
or a constituent quark mass of $m_q=\frac{1}{3}m_N$. Thus  and $b_N=0.365$F we obtain $\hbar \omega_q=293$ MeV.

{\it Model B}:
A more realistic case would be to consider a harmonic confining potential between quarks of the form:
\beq
V_c=-C(\lambda,\mu)\frac{1}{2}\sum_{i<j}^{3}k ({\bf r}_i-){\bf r}_j)^2=-C(\lambda,\mu)\frac{1}{2} \omega_q^2 \sum_{i<j}^{3}({\bf r}_i-){\bf r}_j)^2
\eeq
We will evaluate the matrix element of of this interaction between the nucleon state and demand that this equal to the rest nucleon mass energy $m_Nc^2$.\\
The result is:
\beq
3 m_q c^2=\frac{1}{2}m_q \omega_q^2b_N^2\frac{8}{3}J_{orb}\Rightarrow 3c^2={8}b_N^2 \omega^2_q\Rightarrow(\hbar \omega)_q=\sqrt{\frac{3}{8}}\frac{\hbar c}{b_N}
\label{Eq:homega3q}
\eeq
Thus for  $b_N=0.65$ F we find 
\beq
(\hbar \omega)_q\approx 188 \mbox{MeV}, 
\eeq
smaller than that  given by Model A, Eq. (\ref{Eq:Homegacon}).

Similarly we find $\langle {\bf R}^2\rangle=\int (0s(R))^2 R^2 dR=(3/2)b_N^2$ or $b_N=\sqrt{2/3} R_N $ which is the standard expression.
%somewhat smaller  the simple result obtained above.

\begin{figure}
	%	\centering
	\begin{center}
		{\includegraphics[width=0.9\textwidth]{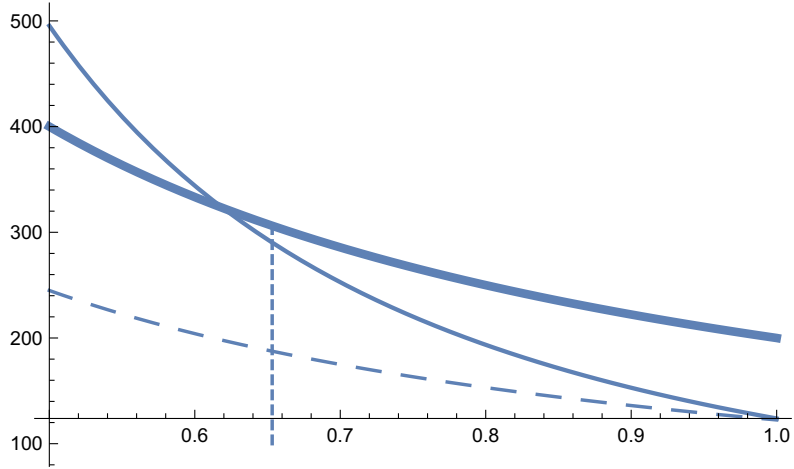}
		}
		\\
		%{\hspace{-1.5cm} $E_R\rightarrow$keV}
		\caption{ Shown are i) the energy scale sc (thick solid line) ii)  $(\hbar  \omega)_q$ (solid line) in model A (constituent quarks) and iii) $(\hbar  \omega)_q$ (dashed line) in model B (confining potential),  all in  MeV, vs the size parameter $b_N$ in F. The values of these parameters corresponding to  a nucleon radius of 0.8 F are indicated by the intersection of these curves with the dotted line, i.e an energy scale of 310 MeV and  $(\hbar  \omega)_q$=293 MeV (Model A) and 190 MeV (Model B)}
		\label{Fig:homegabN}
	\end{center}
\end{figure}

The above formalism allows one to simply estimate the binding energy of the nucleon using the one gluon exchange potential.
% Now the radial integral 
%takes the form:
%$$J_{orb}=\int|0s(\xibf)|^2|0s(\etabf)|^2d^3\xibf\left (\frac{1}{\sqrt{2}|\xibf|}+\frac{1}{\sqrt{2}|\frac{1}{2}\xibf+\frac{\sqrt{3}}{2}\etabf|}+\frac{1}{\sqrt{2}|-\frac{1}{2}\xibf+\frac{\sqrt{3}}{2}\etabf|}\right )=3 \sqrt{2}\frac{1}{\sqrt{\pi}}$$
%This result is obtained  by performing  an orthogonal transformation, ${\bf u}=\frac{1}{2}\xibf+\frac{\sqrt{3}}{2}\etabf,\,{\bf v}=-\frac{1}{2}\xibf+\frac{\sqrt{3}}{2}\etabf$. Then one finds:
%$$\int|0s(\xibf)|^2|0s(\etabf)|^2d^3\xibf d^3\etabf\left (\frac{1}{\sqrt{2}|\frac{1}{2}\xibf+\frac{\sqrt{3}}{2}\etabf|}+\frac{1}{\sqrt{2}|-\frac{1}{2}\xibf+\frac{\sqrt{3}}{2}\etabf|}
%\right )$$ $$ =\int|0s({\bf u})|^2|0s({\bf v})|^2d^3{\bf u}d^3{\bf v}\left (\frac{1}{\sqrt{2}|{\bf u}|}+\frac{1}{\sqrt{2}|{\bf v}|}\right )$$
%The last integrals are of the standard form.\\
The total Matrix element is:
\beq
E_b=200 \mbox{MeV}  (-\frac{8}{3})3 \sqrt{2}\frac{1}{\sqrt{\pi}} =-1260\mbox{MeV}
\eeq
The six quark cluster is expected to  about twice that.

%%%%%%%%%%%%%%%%%%%%%%%%%%%%%%%%%%%%%%%%%%%%%%%%%%%%%%%%%%%%%%%%%

\subsection{Multi-Quark Matrix Elements}

The procedure is straightforward.  The interaction $V( |r_1 - r_2|)$ is diagonalized in the basis defined above.  In this paper only a central interaction is used so there are no non-zero matrix elements between states of different $L$ or $S$.  A non-central interaction would produce admixtures, but in the absence of any indication that such interactions are required at the quark-quark, we stay with the simpler choice.
 The $2\hbar\omega$ levels are given an additional diagonal energy of $\Delta \approx$ 300-500 Mev.

The matrix element $<0s^6\;  |V|\, 0s^6 >$ is the simplest case of multi-quark matrix elements but the techniques used are similar to the remaining cases.  Two-body matrix elements for $0s^5$ and
$0s^4$ will be similar except for the labels and the tedium \cite{Str-Ver}.    
We take first the $L=S=0$ case.

The matrix elements of a two-body operator are then:

\begin{eqnarray}
&&\hspace{-.9cm}< (0s)^6\, [6]_r(0 0)_r\; [2^21^2]_{cs}(0 0)_c S=0\Big|\;\sum_{i,j}V_{ij}\;\Big|
(0s)^6\, [6]_r\,(0 0)_r\; [2^21^2]_{cs}(0 0)_c S=0 >
\nonumber\\
&=&\frac{7}{2}<(2\; 0)_{c}\;S_{2}=0\;I_2=1>+
\frac{13}{2}<(0\;1)_{c}\;S_{2}=1\;I_2=1>+\nonumber\\
&&\frac{5}{2}<(2\;
0)_{c}\;S_{2}=1\;I_2=0>+\frac{5}{2}<(0\; 1)_{c}\;S_{2}=0\;I_2=0>\;\;\;\;\;\;\;\;\label{eq:first1}\\
&\rightarrow& - <\;S_{2}=0\,> - 7 <\;S_{2}=1\,>     \nonumber\label{eq:first1}
\end{eqnarray}

where 
\begin{eqnarray}
<(\lambda_2 \mu_2)_{c}\;S_{2}>=  \Big< (0s)^2\, \;
[f_2]_{cs}(\lambda_2 \mu_2 )_c \;S_2 \;I_2 \;J_2\Big| \;V_{56} \;\Big|
(0s)^2\, \; [f_2]_{cs}(\lambda_2 \mu_2)_c \;S_2 \;I_2\;J_2
\Big>\nonumber
\end{eqnarray}
and the expression to the right of the arrow results from inserting the expectation values of $\lambda_1\cdot\lambda_2$ and assuming no explicit isospin dependence in the interaction.  Thus, for a spin-independent interaction, the energy of $0s^6$ is just a multiple of the energies of $0s^2$ states.

%Formulae for  the  other matrix elements may be found  in our previous paper \cite{Str-Ver}. 

\subsection{Removal of Spurious States}

Diagonalization of the interactions produce wave functions that can be used to interact with the remaining $A-2$ nucleons.   However, the resulting wave functions contain components from spurious states.  As first pointed out by Skyrme\cite{Skyrme}, when converted to center-of-mass and relative coordinates, the $2\hbar\omega$ wave functions contain parts in which the center-of-mass is not in the ground state.  Including such components can modify calculated transition rates and energies.  Therefore we project out  such spurious states.  For $J=0, I=1$ there are seven singly spurious states and two doubly spurious states.

An elegant method using $SU(3)$ of projecting spurious states from the resulting wave functions was subsequently developed by Hecht\cite{spur}.  It relies on the fact that the centre-of-mass operator $\bf R$ transforms under $SU(3)$ as $(1 0)$.  Thus, operating on an $SU(3)$ state $(\lambda \mu)$ results in spurious states with $SU(3)$ representation $(\lambda +1\: \mu), (\lambda -1\: \mu +1)$, or $(\lambda \:\mu -1)$.  There will normally be other states with these representation, but with Hecht's method  one can more easily remove the spurious components.  Hecht's method is equivalent to the one used in this paper.

\subsection{The Structure of the $q^6$ Eigenstates
\label{sec:MulttqME}} 

We shall focus on the $I=1, J=0$  six-quark system which is most relevant below.  Results for three-quark and other six-quark states, both positive and negative parity, will be published elsewhere.  The calculations assumed $\alpha_s$= 1 and $b_q = 0.8 $ fm. The energies of the 2$\hbar\omega$ states were increased by $\Delta$ = 300-500 MeV.  The spurious states were projected out in all cases.  

The QCD interaction produces relatively little mixing between the ground state and the 2$\hbar\omega$ states.  This is in part because the lowest lying 2$\hbar\omega$ state before projection is largely spurious.  After projection the ground state is 99\% $0s^6$.  As  already stated the $0s^6\;J=1, I=0$ state and the  $0s^6\;J=0, I=1$ states are degenerate using a spin-independent interaction.  This degeneracy is broken by the effects of the 2$\hbar\omega$ states.

The Ohta interaction was fit somewhat to experiment and one can expect that its results reflect this to some extent.  The $I=0, J=1$ state is separated from the $I=1, J=0$ state by over 70 MeV.  The  $I=1, J=0$ ground state is 93\% $0s^6$ for $\Delta$ = 500 MeV and its energy is depressed by 71 MeV by the effects of the 2$\hbar\omega$ states.  The first excited  $I=1, J=0$ level - which could help determine $\Delta$ - lies 423 MeV higher, although such a state could also be explained by $q^8 \bar{q}^2$ states.

\section{Separation into Two Baryon Clusters, $q^3 \times q^3$}
\label{sec:BaryonClusters}

\begin{table}\centering
	\begin{tabular}{|c|c|c||c|c|c|c|c|c|c|cl}
		\hline
		$0s^6$ &$0s^5$ &$0s^4$ &  L & S & J & I &singly&doubly\\\hline
		\hline
		1 & 2 & 7 & 0 & 1 & 1 & 0 & 1 & 1 \\
		0 & 0 & 4 &1 & 0 & 1 & 0 & 1 & 0 \\
		0 & 0 & 3 & 1 & 1 & 1 & 0 & 1 & 0 \\
		0 & 2 & 7 & 2 & 1 & 1 & 0 & 1 & 1 \\
		0 & 0 & 4 & 1 & 2 & 1 & 0 & 2 & 0 \\
		0 & 2 & 3 & 2 & 2 & 1 & 0 & 2 & 0\\
		0 & 1 & 2 & 2 & 3 & 1 & 0  & 0 & 1\\ \hline\hline
		1 & 2 & 7 & 0 & 0 & 0 & 1 & 1 & 1\\
		0 & 0 & 11 & 1 & 1 & 0 & 1 & 4 & 0\\
		0 & 3 & 7 & 2 & 2 & 0 & 1 & 2 & 1\\
		\hline
	\end{tabular}
	\caption[Basis States]{Six-quark states, even parity, of relevance to the current calculations.  The three left-most columns specify the number of basis states with $0s^n$ for a given $L,S,J,I$, the  next four columns give the number of states for a given $L,S,J,I$.  The last two columns give the number of singly or doubly spurious states for a given $L,S,J,I$.  {\it E.g.} for $L=0=J=0, I=1$, there are 10 basis states of which nine are $2\hbar\omega$ excitations; one combination of the states is singly spurious and one is doubly spurious.} \label{tab:6qbasis}
\end{table}

To investigate the effect of three-quark clusters on electroweak transitions it is necessary to convert the six-quark states into two three-quark clusters.  One can do this in two ways.  The methods produce identical results.

\subsection{Separation Using $SU(6)_{cs} \subset SU(3)_c\times SU(2)_s$ Algebra}

In the first method one uses the recoupling coefficients and algebra of $SU(6)_{cs} \subset SU(3 )_c\times SU(2)_s$.
  Here, we briefly explain the formulae required using schematic formulae.

Let $K$ represent the variables $(\lambda \mu) S, I$ for the six-quark state and $k$ for a single quark; the significance of the other $K$ variables will be obvious..  Then, one has

\barr
| q^6 K >&=& \sum_{K_3, K_{23},K_5} < q^5 K_5 \times q \;k ;K\; |\} q^6 \;K> < q^3 K_3\times q^2 \;K_{23}; K_5 |\} q^5 \;K_5 >  \nonumber\\
&\times& \Big| \; \Big[ [\; |\;q^3 K_3\times q^2K_{23} > ]^{(K_5)} \times q \Big]^{(K)} >
\earr
where
\barr
&&\Big| \; \Big[ [\; |\;q^3 K_3\times q^2K_{23} > ]^{(K_5)} \times q \Big]^{(K)} >=\sum_{K'} \times q\:k >  ]^{(K')} \Big]^{(K)} >\nonumber  \\
&&|\: q^2  K_{23}  \times q\:k ;\;K'  >   =   \sum  < q^3 K' \;|\: q^2  K_{23}  \times q\:k ;\;K'  > | \; q^3 K'  \:> .\nonumber\\
\earr

The Racah coefficient $U(K_3\: K_{23}\: K\: k; K_5\: K') $ in reality is the product of an SU(3) Racah coefficient and two SU(2) Racah coefficients, one for spin and one for isospin.  The factor $< q^5 K_5 \times q \;k ;K\; |\} q^6 \;K>$ was called a fractional parentage coefficient by Racah\cite{Racah2, Talmi} and can be written as a product of $SU(6_{cs} ) \subset SU(3)_c\times SU(2)_\sigma$ coefficients.
Note that the six-quark cluster is a color singlet.  When one separates the six-quarks into two three-quark entities, the individual three-quark cluster can be in a colour singlet or  a colour octet.   

Separation for the 2$\hbar\omega$ states proceeds similarly.  We omit the equations here.

%%%%%%%%%%%%%%%%%%%%%%%%%%%%%%%%%%%%%%%%%%%%%%%%%%%%%%%%%%%%%%%%%%%%%%%%%%%%%%%%%%%%%%%%
\subsection{Orbital expansion}
In the six-quark cluster we encounter the following orbital combinations:\\
i) $0s^{6}$ states:
$$\Big|  [0s(\bf{r}_1) 0s(\bf{r}_2) 0s(\bf{r}_3)][0s(\bf{r}_4) 0s(\bf{r}_5) 0s(\bf{r}_6)] \;\big >$$
ii) $0s^{5}1s$ states:
$$ \Big|  [0s(\bf{r}_1) 0s(\bf{r}_2) 0s(\bf{r}_3)][0s(\bf{r}_4) 0s(\bf{r}_5) 1s(\bf{r}_6)] \;\big >$$
iii) $0s^{5}0d$ states:
$$ \Big|  [0s(\bf{r}_1) 0s(\bf{r}_2) 0s(\bf{r}_3)][0s(\bf{r}_4) 0s(\bf{r}_5) 0d(\bf{r}_6)] \;\big >$$
iv)  $0s^{4}0p^2L$ states, $L=0,1,2$:
$$ \Big|  [0s(\bf{r}_1) 0s(\bf{r}_2) 0s(\bf{r}_3)][0s(\bf{r}_4)  0p(\bf{r}_5)\times0p(\bf{r}_6)L] \;\big > .$$

We know that
$$ \Big|  [0p(\bf{r}_5)\times0p(\bf{r}_6)L=0]=\frac{1}{\sqrt{2}} \Big[ | 0s\left (\frac{\bf{r}_5-\bf{r}_6}{\sqrt{2}}\right )1s\left (\frac{\bf{r}_5+\bf{r}_6}{\sqrt{2}}\right ) >$$ 
$$ + | 1s\left (\frac{\bf{r}_5-\bf{r}_6}{\sqrt{2}}\right )0s\left (\frac{\bf{r}_5+\bf{r}_6}{\sqrt{2}}\right )> \Big]$$

$$\Big|   [0p(\bf{r}_5)\times0p(\bf{r}_6)L=2 >=\frac{1}{\sqrt{2}} \Big[ | 0s\left (\frac{\bf{r}_5-\bf{r}_6}{\sqrt{2}}\right )0d\left (\frac{\bf{r}_5+\bf{r}_6}{\sqrt{2}}\right )>$$ $$- | 0d\left (\frac{\bf{r}_5-\bf{r}_6}{\sqrt{2}}\right )0s\left (\frac{\bf{r}_5+\bf{r}_6}{\sqrt{2}}\right )> \Big]$$
$$ \Big|  [0p(\bf{r}_5)\times0p(\bf{r}_6)L=1]=[0p(\bf{r}_5)\otimes0p(\bf{r}_6)]$$
We wish to express them in terms of the relative coordinates of the three-quark cluster defined by
\beq
\xibf=\frac{1}{\sqrt{2}}({\bf r}_4-{\bf r}_5),\,\etabf=\frac{1}{\sqrt{6}}({\bf r}_4+{\bf r}_5-2 {\bf r}_6),\,{ \bf R}=\frac{1}{\sqrt{3}}({\bf r}_4+{\bf r}_5+{ \bf r}_6)
\eeq
The inverse transformation
$$
\left ( \begin{array}{c} {\bf r}_4\\{\bf r}_5\\{\bf r}_6\\ \end{array} \right )=\left ( \begin{array}{rrr} \frac{1}{\sqrt{2}}& \frac{1}{\sqrt{6}}& \frac{1}{\sqrt{3}}\\-\frac{1}{\sqrt{2}}& \frac{1}{\sqrt{6}}&\frac{1}{\sqrt{3}}\\ 0&- \frac{2}{\sqrt{6}}& \frac{1}{\sqrt{3}}\\ \end{array} \right )\left ( \begin{array}{c} \xibf\\\etabf\\{\bf R}\\ \end{array} \right )
$$
is useful.\\
For the cluster of the first three-quarks this is trivial:
$$ [0s(\bf{r}_1) 0s(\bf{r}_2) 0s(\bf{r}_3)]=\psi(\xibf_1,\etabf_1)0s({ \bf R}_1)$$
where $\psi(\xibf_1,\etabf_1) $ is the relative three-quark wave function of the form $ 0s(\xibf_1),0s(\etabf_1)$. 
Similarly
$$ [0s(\bf{r}_4) 0s(\bf{r}_5) 0s(\bf{r}_6)]=\psi(\xibf_2,\etabf_2)0s({ \bf R}_2) .$$
Now
$$ [0s(\bf{r}_4) 0s(\bf{r}_5) 1s(\bf{r}_6)]\propto (\frac{3}{2}-{\bf r}^2_6)=\frac{1}{3}\left (\frac{3}{2}-{\bf R}_2^2\right ) +\frac{2}{3}\left (\frac{3}{2}-\etabf _2^2\right )+\frac{\sqrt{2}}{3}\etabf_2\cdot {\bf R}_2$$

The interesting term is the first in the sum (we do not want excitations in the relative coordinates). The Gaussian exponent remains invariant under the change of coordinates. Thus
$$[0s(\bf{r}_4) 0s(\bf{r}_5) 1s(\bf{r}_6)]=\frac{1}{3}\psi(\xibf_2,\etabf_2)1s({ \bf R}_2) + \cdots$$ 
In an analogous fashion we find
$$[0s(\bf{r}_4) 0s(\bf{r}_5) 0d(\bf{r}_6)]=\frac{1}{3}\psi(\xibf_2,\etabf_2) 0d({ \bf R}_2)+ \cdots $$
The $ | 0p^2L=1 >$ state cannot be made to be of the form of the relative function $\psi(\xibf_2,\etabf_2)$ so it does not make a contribution.
$$ \langle | 0s(\bf{r}_4)]0p^2L=1 >\rangle=0$$

The state  $ \ 0p^2L=0 >$ can be expressed in terms of $0s$ and $1s$ states. Similarly the $| 0p^2 L=2>$ can be expressed in terms of $0s$ and $0d$ states. Proceeding as above we get:
$$ \langle 0s(\bf{r}_4)]0p^2L=0\rangle=\frac{1}{3\sqrt{2}}\psi(\xibf_2,\etabf_2)1s({ \bf R}_2)$$
$$ \langle 0s(\bf{r}_4)]0p^2L=2\rangle=\frac{1}{3\sqrt{2}}\psi(\xibf_2,\etabf_2)0d({ \bf R}_2)$$

We should mention that in the L expansion one could use another scheme in the case of the $0p^2$ configurations  another possible coupling involving involving the product is possible:
$0s({\bf r})_10s({\bf r}_2)0p({\bf r}_3)$  is transformed as follows
$$0s({\bf r})_10s({\bf r}_2)0p({\bf r}_3)=-\frac{2}{\sqrt{6}}\psi(\xibf_1),0p(\etabf_1)0s({ \bf R}_1)+\frac{1}{\sqrt{3}}\psi(\xibf_1),0s(\etabf_1)0p({ \bf R}_1) .$$
Thus the orbital part relevant for the 2-nucleon like components is
\beq
|(0p^2)L\rangle =\frac{1}{3}\psi^2(\xibf,\etabf)|(0p^2({\bf R})L\rangle .
\eeq
For $L=0,2$ we find the presence of the same components as above, while  the $L=1$ is new. Depending on  the appropriate color spin-isospin symmetries, consistent with orbital symmetry $ 0s^2[2]0p\times 0s^2[2]0p$, these contributions may be important.

We note  that the coordinate $ {\bf R}$ represents the center of the three-quark cluster which can be identified with the coordinate of the nucleon. In our case this is a harmonic oscillator wave function defined with $\hbar \omega_q$.

\begin{table}[h]\centering
	\begin{tabular}{|c|c|c|c|c|c|c|c|c|c|c|c||} 
		\hline
		$0s^6$ & $[f]_{cs}$ & $(\lambda,\mu)$ &&&& &$S$ & $L$ & $\kappa_{orb}\times\kappa^{0s^6}_{NN }$  \\\hline
		1 & $[2^21^2]$ & $(0,0)$ && &&& 0 & 0 & $1/3 $\\		
		\hline\hline
		$0s^51s $ & $[f]_{cs} $& $(\lambda_1,\mu_1) $ & $ S_1I_1$&$[f_2]$ & $(\lambda_2,\mu_2$ & $S_2$  $I_2$ & $S $& $L$ & $\kappa_{orb}\times \kappa^{0s^51s}_{NN} $ \\\hline
		$	2 $ & $[2^21]$ & $(0,1)$ & $\frac{1}{2} \frac{1}{2} $& $[1]$ & $(1,0)$ & $\frac{1}{2} \frac{1}{2} $  & 0 &  0& $5/27$\\
		$3 $ & $[21^3]$ & $(0,1)$ & $\frac{1}{2} \frac{1}{2} $& $[1]$ & $(1,0)$ & $\frac{1}{2} \frac{1}{2} $ & 0 & 0 & $-4/27$\\							\hline\hline				
		$0s^4 0p^2$ & $[f]_{cs} $&$ (\lambda_1,\mu_1)$ & $S_1 I_1$ & $[ f_2]$ & $(\lambda_2,\mu_2)$ & $S_2$  $I_2$ & $S$ & $L$ & $\kappa_{orb}\times \kappa^{0s^40p^2}_{NN}$  \\\hline
		4 &$[2^2]$& (1,0) &1 0& $[1^2]$ &(0,1)&1 1&0&0&$1/(9\sqrt{3})$ \\
		5& $[21^2]$     &(1,0)&1 1& $[1^2]$ &(0,1)&1 1&0&0&$1/ (9\sqrt{2})$ \\
		6& $[21^2]$   &(1,0)&0 1&[2]&(0,1)&0 0&0&0&$1/(6\sqrt{6})$  \\
		\hline
		7 &$[2^2]$&(1,0)&1 0&[2]&(0,1)&0 1&1&1&0  \\
		8&$[2,1^2]$ &(1,0)&1 1&[2]&(0,1)&0 1&1&1&0  \\
		9&$[2,1^2]$ &(1,0)&0 1& $[1^2]$ &(0,1)&1 0&1&1&0  \\
		10&$[2,1^2]$ &(1,0)&1 1& $[1^2]$ &(0,1)&1 0&1&1&0 \\	
		\hline
	\end{tabular}
	\caption{The six-quark $I=1,\,J=0$  basis  states and the non-zero spin-isospin  coefficients $\kappa_{orb}\times \kappa$, needed to transform them into two nucleon-like configurations. }
	\label{tab:taba}
\end{table}

% ii) {$J=1,I=0$ states}\\
\begin{table}[h]\centering
	\begin{tabular}{|c|c|c|c|c|c|c|c|c|c|c|c|}
		\hline
		$0s^6$ & $[f]_{cs}$ & $(\lambda,\mu)$ &&&&& $S$ & $L$ & $\kappa_{orb}\times \kappa^{0s^6}_{NN }$  \\\hline
		1 &$[2^3]$&(0,0)&&&&&1&0&$\sqrt{5}/6$\\\hline\hline
		$0s^51s $ & $[f]_{cs} $& $(\lambda_1,\mu_1) $ & $ S_1 I_1$& $[f_2]$ & $(\lambda_2,\mu_2)$ & $S_2  I_2$& $S $& $L$ & $\kappa_{orb}\times \kappa^{0s^51s}_{NN} $ \\\hline
		2 & $[2^21]$&(0,1)&$\frac{1}{2} \frac{1}{2} $&[1]&(1,0)&$\frac{1}{2} \frac{1}{2} $&1&0&5/27\\
		3 & $[2^21]$&(0,1)&$\frac{3}{2} \frac{1}{2} $&[1]&(1,0)&$\frac{1}{2} \frac{1}{2} $&1&0&$2\sqrt{5}/27$\\
		\hline
		$0s^4 0p^2$ & $[f]_{cs} $&$ (\lambda_1,\mu_1)$ & $S_1 I_1$ & $[ f _2]$ & $(\lambda_2,\mu_2)$ & $S_2 I_2$ & $S$ & $L$ & $\kappa_{orb}\times \kappa^{0s^40p^2}_{NN}$  \\\hline
		4 & $[2^2]$ &(1,0)&1 0&[2]&(0,1)&0 0&1&0&-$1/6$  \\
		5 & $[2,1^2]$ &(1,0)&1 1& $[1^2]$ &(0,1)&1 1&1&0&$\sqrt{3}/18$  \\
		
		6 & $[2,1^2]$ &(1,0)&1 1& $[1^2]$ &(0,1)&1 1&1&0&$1/(9\sqrt{2})$ \\
		7 & $[2^2]$ &(1,0)&1 0&[2]&(0,1)&0 1&1&2&0 \\
		%\hline	
		8 & $[2,1^2]$ &(1,0)&0 1& $[1^2]$ &(0,1)&1 1&1&2&0  \\
		9     & $[2,1^2]$ &(1,0)&2 1& $[1^2]$ &(0,1)&1 1&3&2&0  \\
		%\hline
		10&   $[2^2]$ 	&(1,0)&1 0& $[1^2]$ &(0,1)&1 0&0&1&0  \\
		11 & $[2^2]$ 	&(1,0)&1 0& $[1^2]$ &(0,1)&0 0&1&1&0 \\
		12&   $[2,1^2]$ 	&(1,0)&0 1&[2]&(0,1)&0 1&0&1&0  \\
		13&   $[2,1^2]$ 	&(1,0)&1 1&[2]&(0,1)&0 1&1&1&0  \\
		\hline
		$0s^50d $ & $[f]_{cs} $& $(\lambda_1,\mu_1) $ & $ S_1 I_1$&$[f_2]$ & $(\lambda_2,\mu_2)$ & $S_2   I_2$& $S $& $L$ & $\kappa_{orb}\times \kappa^{0s^50d}_{NN} $ \\\hline
		14 & $[2^21]$ &(0,1)&$\frac{1}{2} \frac{1}{2} $&[1]&(1,0)&$\frac{1}{2} \frac{1}{2} $&1&2&0\\
		15 & $[2^21]$ &(0,1)&$\frac{3}{2} \frac{1}{2} $&[1]&(1,0)&$\frac{1}{2} \frac{1}{2} $&1&2&0\\
		\hline
	\end{tabular} 
	\caption{The six-quark  $I=0,\,J=1$  basis states and the non-zero spin-isospin coefficients $\kappa_{orb}\times\kappa$ needed to transform them into two nucleon-like configurations.}
	\label{tab:tabb}
\end{table} 

\subsection{Color Spin-Isospin Projection}

The possible color-spin three-quark configurations for each isospin  are:
\barr
&&[21]_{cs}:(30)\frac{1}{2},\,(11)\frac{1}{2},\,(11)\frac{3}{2},\,(00)\frac{1}{2} \;\;\;{\rm for}\;\;\;  I=\frac{1}{2}\;\;\mbox{ and }\nonumber\\&&[1^3]_{cs}:(11)\frac{1}{2},\,(00)\frac{1}{2} \;\hspace{2.4cm}{\rm for}\;\;\; I=\frac{3}{2} .
\earr
For the $I=0$ $S=1$ case we will consider all such configurations. In the case of the $I=1$ $S=0$ we will limit ourselves  to expressing the six-quark wave functions as a product of two three-quark color-singlet clusters with the spin-isospin quantum numbers of  the nucleon, {\it e.g.},
\beq
0s^6_{IS}=\kappa^{0s^6}_{IS}\left(\left[2,1\right](00)\frac{1}{2}\times \left[2,1\right](00)\frac{1}{2} \right)_{IS}
\eeq

The required  transformation coefficients $\kappa$ are evaluated using the formalism of section \ref{sec:BaryonClusters} based on the material found in  appendix D, section \ref{AppendixD}. We note that they vanish for states with $S >1$ and  in the case of  $0p^2$ states with color-spin symmetry $(01)_c 0$ and $(01)_c1$ can lead to two-nucleon like configurations, but the $L=1$ states cannot, due to orbital selection rules. 

The results for for the coefficients of the color singlet [21] three-quark states are shown    in tables \ref{tab:taba} and \ref{tab:tabb} in the basis indicated there as $\kappa[1],\kappa[2]$ {\it etc.}

For orientation purposes  sometimes the $\kappa$'s may carry appropriate indices, {\it e.g.} for the $0p^2$ states  using the notation $\kappa^{0s^40p^2}_ { NN}$, 
%i) {$J=0,I=1$ states}\\
%Total number of $I=1,J=0$ states:1+2+4+24=31 (see Table \ref{tab:taba})\\
%Total number of $I=0,J=1$ states:1+2+5+30=38 (see Table \ref{tab:tabb})

\subsection{Combined Orbital-Color Spin-Isospin Projection}
\label{sec:snug}
we have seen above that the orbital part is of the projection is given by
\barr
0s^6&=&(0s^2)^2 0s({\bf R}_1)0s({\bf R}_2),0s^51s=\frac{1}{3}(0s^2)^2 0s({\bf R}_1)1s({\bf R}_2),\nonumber\\
0s^4(0p^2L=0)&=&\frac{1}{3\sqrt{2}} (0s^2)^2 0s({\bf R}_1)1s({\bf R}_2)
\earr
The part $0s^2$ refers to the internal structure of each nucleon-like component, which is of no particular interest, and for convenience it will be subsequently  omitted. All other orbital configurations do not lead to two nucleon like configurations, since hey are expected to occur in the ground state of the relevant $6-q$ cluster. Thus we can write  a factor $\kappa_{orb}=1,1/3,1/(3\sqrt{2})$ respectively for the above configurations. For all ather configurations is taken to be zero.\\
The  color spin isospin part depends, of course, on the configuration and be calculated in a similar fashion as above so we can write:
\beq
|i\rangle=\kappa_{0rb}(i)\kappa(i) ns n's\left(\left[2,1\right](00)\frac{1}{2}\times \left[2,1\right](00)\frac{1}{2} \right)_{IS},nsn's=\left \{ 
\begin{array}{cc} 0s 0s, i=1\\
	0s 1s, \mbox{otherwise}\\
	%	\right .
\end{array}
\right .
\eeq

The interesting conclusion that in fact we encounter only two nucleon like configurations for given spin and isospin.
thus we find  
\beq
\Phi(6q; J,I)=A_{0s}[J,I]\,\left [0s\frac{1}{2}\times 0s\frac{1}{2}\right ]^{J,I}+A_{1s}[J,I]\,\left [0s\frac{1}{2} \times 1s\frac{1}{2}\right ]^{J,I}
\eeq
where
$$ A_{0s}[J,I]=(c[1] \kappa[i])_{J,I},\,A_{1s}=\sum_{i=2} (c[i]\kappa_{orb}[i]\kappa[i])_{J,I}$$
where c[i] is the amplitude of  the   configuration $i$  in the $\Phi(6q; J,I)$ (for the relevant values  see tables \ref{tab:taba} and \ref{tab:tabb}). The values of $c[i]$ will be given below.

i) $J=0, I=1$ case.\\
We find that the coefficients are:
$$ c=\{0.993, 0.051, -0.053, 0.052, 0.053, 0.051\}$$
and
\beq
A_{0s}[0,1]=0.331,\,A_{1s}[0,1]=0.027
\eeq
ii) $J=1, I=0$ case.\\
Now the coefficients are:
$$ c=\{0.994, 0.049, 0.048, -0.050, 0.048, 0.049\}$$
and
\beq
A_{0s}[1,0]=0.907,\,A_{1s}[1,0]=0.035
\eeq
%.taton  involving the groups $SU(6)_{cs}$ and$ SU(2)_I$ can be evaluated using the techniques of section \ref{Eq:separtionscI}.

There remains another task. In order to evaluate the nuclear matrix elements  coupling the $6-q$ cluster to nuclei, as discussed in section \ref{sec:MicBSM}. one must now expand the radial wave functions from $(\hbar \omega)_q$  to the $(\hbar \omega)_N$ employed in the nuclear shell model.
Thus,
\beq
\big | 0s({ \bf r}_1)0s({ \bf r}_2) > =\sum_{n1\le n_2} C^{0s,0s}_{n_1n_2} \;\big | (n_1,0)(n_2,0)L=0\rangle,\,C^{0s,0s}_{n_1n_2}=\langle n_1,0|0s\rangle\langle n_2,0|0s\rangle .\nonumber
\eeq
with similar equations for the other two-quark states.  
The expansion does not converge very fast, but only a few terms in the sum are adequate for our purposes.  Many radial components appear:
\barr
|f_{1,0}\rangle&=&\sum_{n1,n2} \left (A_1|C^{0s0s}_{n_1,n_2}+A_2|C^{0s1s}_{n_1,n_2}\right )\nonumber\\ 
&&\times |(n_1,\ell_1=0),(n_2,\ell_2=0) ,L=0,S=0,J=0,I=1 \rangle 
\label{Eq:expn1n2}
\earr

The  coefficients $C^{0s0s}_{n_1,n_2}$  are given in table \ref{tab:n1n2J0} in the appendix
for $b_A/b_q=1.5.\,2.0,\,2.8$, with $b_A$ being the harmonic oscillator length for the nucleon in the nucleus and $b_q$ that of the quarks in the six-quark cluster. Theses values of $b_A/b_q$  are relevant for nuclei with  mass from 2 to 48. 

Before concluding the section it is worth mentioning that components other than the two-nucleon ones can, of course,  be examined as members of the cluster including colored ones. Thus, {\it e.g.}, in the case of the dominant  $0s^6$ $I=0,\,S=1$ configuration  we obtain the  expansion with the three-body coefficients found in table \ref{tab:exotic01}.

\begin{table}[h]\centering
	\begin{tabular}{|ccc||c|c|c|c|c|c|c|cl}
		\hline
		$q^3[f_1](\lambda_1\mu_1)S_1\,I_1$ & $ \times$ & $ q^3[f_2](\lambda_2\mu_2)S_2\,I_2 $ & cfp\\\hline
		$[21](11)\frac{1}{2}\frac{1}{2}$ & $\times $ & $[21](11)\frac{1}{2}\frac{1}{2}$ & $-\frac{\sqrt{5}}{3\sqrt{2}}$\\
		$[21] (11)\frac{1}{2}\frac{1}{2}$ & $\times $ & $[21](11)\frac{3}{2}\frac{1}{2}$ & $-\frac{\sqrt{5}}{3\sqrt{2}}$\\
		$[21] (11)\frac{3}{2}\frac{1}{2}$ & $\times $ & $[21](11)\frac{1}{2}\frac{1}{2}$ & $\frac{\sqrt{5}}{3\sqrt{2}}$\\
		$[21] ](11)\frac{3}{2}\frac{1}{2}$ & $\times $ & $[21](11)\frac{3}{2}\frac{1}{2}$ & $\frac{\sqrt{5}}{6}$ \\
		$[21] (00)\frac{1}{2}\frac{1}{2}$ & $\times $ &$ [21](00)\frac{1}{2}\frac{1}{2}$ & $-\frac{1}{6}$ \\
		$[111](11)\frac{1}{2}\frac{3}{2}$ & $\times $ & $[1^3](11)\frac{1}{2}\frac{3}{2}$ & $\frac{\sqrt{5}}{3}$ \\
		$[111](00)\frac{3}{2}\frac{3}{2}$ & $\times $ & $[1^3](00)\frac{3}{2}\frac{3}{2}$ & $-\frac{2}{3}$ \\
		\hline
	\end{tabular}
	\caption[Three-body cfps, I=0, J=1]{Three-body cfps, T=0, J=1\label{tab:exotic01}}
\end{table}

%\clearpage

This can of course be a component of the deuterium. Furthermore  it  is claimed \cite{DiBarNuc20} that this provide  a strong enrichment of the high-momentum component for $^6$Li considered in  $\alpha+0s^6$ (I=0,J=1) cluster model. Note that the color singlet component occurs with  about 47$\%$ probability. Hadron components other than the nucleon are also present.  
For the $I=1,J=0$ case, table \ref{tab:exotic10}, components other than two nucleon-like clusters are possible.  

\begin{table}\centering
	\begin{tabular}{|ccc|c|c|c|c|c|c|c|cl}
		\hline
		$q^3[f_1](\lambda_1\mu_1)S_1\,I_1$ & $\hspace{-.4cm}\times \hspace{-.2cm}$ & $q^3[f_2](\lambda_2\mu_2)S_2\,I_2$ & $cfp$ \\\hline
		$[21] (11)\frac{1}{2}\frac{1}{2}$ & $\times $ & $[21](11)\frac{1}{2}\frac{1}{2}$ & $\frac{\sqrt{2}}{3}$ \\
		$[21] (11)\frac{1}{2}\frac{1}{2}$ & $\times $ & $[21](11)\frac{3}{2}\frac{1}{2}$ & $\frac{1}{3}$ \\
		$[21] (00)\frac{1}{2}\frac{1}{2}$ & $\times $ & $[21](00)\frac{1}{2}\frac{1}{2}$ & $\frac{1}{3}$ \\
		$[21] (11)\frac{1}{2}\frac{1}{2}$ & $\times $ & $[111](11)\frac{1}{2}\frac{3}{2}$ & $\frac{\sqrt{2}}{3}$ \\
		$[111](11)\frac{1}{2}\frac{3}{2}$ & $\times $ & $[21](11) \frac{1}{2}\frac{1}{2}$ & $\frac{\sqrt{2}}{3}$ \\
		$[111] (00)\frac{3}{2}\frac{3}{2}$ & $\times $ & $[1^3](00)\frac{3}{2}\frac{3}{2}$ & $\frac{2}{3 \sqrt{5}}$ \\
		$[111] (11)\frac{1}{2}\frac{3}{2}$ & $\times $ & $[111](11)\frac{1}{2}\frac{3}{2}$ & $\frac{1}{3 \sqrt{5}}$ \\\hline
	\end{tabular}
	\caption[Three-body cfps, I=1, J=0]{Three-body cfps, I=1, J=0\label{tab:exotic10}}
\end{table}
The color singlet component is about 20$\%$, the color singlet  nucleon-like part  is even smaller, 11$\%$.
Again \cite{DiBarNuc20},  this can provide  a strong enrichment of the high-momentum component for $^6$He considered in  $\alpha+q^6 (I=1,J=0)$ cluster model.

%%%%%%%%%%%%%%%%%%%%%%%%%%%%%%%%%%%%%%%%%%%%%%%%%%%%%%%%%%%%%%%%%%%%%%%%%%%%%%%%%%%%%%%
%%%%%%%%%%%%%%%%%%%%%%%%%%%%%%%%%%%%%%%%%%%%%%%%%%%%%%%%%%%
\section{Mixing of the Cluster States with the Two-Particle Shell Model States}
\label{sec:MicBSM}
Once the cluster states are expressed in terms of the shell model, their mixing to the standard shell model states is achieved via the effective two-nucleon interaction. 
We consider the coupling of the six-quark cluster $\Phi(6q; J_c=0,I_c=1,(I_c)_3=\pm1) = | q^6\; 0, 1, \pm1> $  to  nuclear states.  Henceforth, we shall use the latter notation unless there is a chance of confusion.  

The simplest possibility  to consider is a closed-shell  nucleus with two nucleons outside the closed  shell. 

{\bf A=6 nuclei, specifically  $^6$Li}\\
We  consider admixtures of the type  $^6$Li+$C_{1,0}  \big|^4$He $\times\; q^6\, 1,0,0>$. The mixing coefficient is given by
\barr
C_{1_,0}=\frac{1}{D}&&\langle ^4\mbox{He}; \times\; q^6\, 1,0,0 |V| ^6\mbox{Li}\;J=1,I=0\rangle,\nonumber\\D&=&E(^6\mbox{Li}-^4\mbox{He})-(E_{6q} J=1, I=0-2E_N)
\earr
Furthermore, for this relatively simple  nucleus
\barr
C_{1_,0}&=&\frac{1}{D}\; \langle q^6\, 1,0,0 |V| 0p^2,J=1,I=0\rangle\nonumber\\
=\frac{1}{D}\sum_{n_1,n_2}&&\hspace{-.5cm} \Lambda_{n_1,n_2}(1,0)\langle [(n_s\frac{1}{2}) (n_{2s}\frac{1}{2})] J=0,I=1|V|0p^2 ,J=1,I=0\rangle\nonumber
\earr
where in the last step we used the results  of section \ref{sec:snug} and
\beq
\Lambda_{n_1,n_2}(J,I)=A_{0s}(J,I)C^{0s,0s}_{n_1,n_2}+A_{1s}(J,I)C^{0s,1s}_{n_1,n_2}
\eeq

 In shell model calculations one finds\cite{Vergados72}
\barr 
\psi(0p^2,1 0)&=&0.7996\, | 0p_{3/2}^2  \,01>-0.258 |0p_{1/2}^2  \,01 >\nonumber\\
&-& 0.6004 |0p_{3/2}0p_{1/2}  \,01>.\nonumber
\earr
In the $LS$ coupling scheme the component of interest is  large with amplitude $0.93447\,|\, 0p^2 L=0,S=1>$.

In a similar fashion we find the admixture of the type \\ $^6$He+$C_{0,1}\;\big |^4$He $\times \;q^6 0,1,1 >$.
Proceeding as above we find:
\barr
C_{0_,1}&=&\frac{1}{D}< q^6\; 0,1,1|V| 0p^2 ,J=0,I=1\rangle\nonumber\\
&=&\frac{1}{D}\sum_{n_1,n_2} \Lambda_{n_1,n_2}(0,1)\langle [(n_s\frac{1}{2}) (n_2s\frac{1}{2})] 0 1|V|0p^2,\;0 1\rangle
\earr
In shell model calculations one finds
$$\psi(0p^2J=1,I=0)=0.97841\,| 0p_{3/2}^2 \,01>+0.20670\,|0p_{1/2}^2 \,01>,$$ 
Similarly, the  dominant  $LS$  component is  large, $0.9182 | 0p^2 L=0,S=0>$.
Both nuclei are therefore  favorable subjects for the present study.

{\bf 1s0d shell A=18 }
The relevant wave functions for mass 18 are well studied, see {\it e.g.} refs.\cite{Inoue1,DDS-Oxford}.  The eigenvectors for the $^{18}$O ground state in the Elliott SU(3) scheme are
$$0.81\, | (4 0)L=0> - 0.46\, | (0 2) L=0> + 0.36\, |(2 1)L=1>$$
or 87\% L=0.  Further, most interactions and transitions favour proximity of the nucleons involved.  The SU(3) representations $(\lambda 0)$ (which are the leading representations for four or fewer particles) contain all the relative $0s (r_{12})$ components.  Thus, just as the mass six nuclei should be favourable candidates to seek clusters, so are the mass 18 nuclei.  

We consider $^{18}$O with admixtures of $^{16}$O$\,\times\, |q^6 \;0,1,1 >$ and $^{18}$F with $^{16}$O$\,\times\,  |q^6\; 0,1,0 >.$
For the former we find
\barr
C_{0_,1}&=&\frac{1}{D}<q^6 \;0,1,1 >|V|\;1s0d^2,J=0,I=1>\nonumber\\
&=&\frac{1}{D}\sum_{n_1,n_2} \Lambda_{n_1,n_2}(1,0)\langle [(n_s\frac{1}{2}) (n_2s\frac{1}{2})] J=0,I=1|V|1s0d^2,J=0,I=1\rangle\nonumber
\label{Eq:sum01}
\earr
while in the second we get
\barr
C_{1_,0}&=&\frac{1}{D}\ < q^6 \;0,1,1  |V| 1s0d^2 10 1>\nonumber\\
&=&\frac{1}{D}\sum_{n_1,n_2}(0,1) \Lambda_{n_1,n_2}\langle [(n_s\frac{1}{2}) (n_2s\frac{1}{2})] J=0,I=0 |V| 1s0d^2,J=1,I=0\rangle\nonumber
\label{Eq:sum10}
\earr
with the energy denominators of the nuclei and the energy of the appropriate six-quark bag as above.  

The above calculations are straightforward. 
They can be extended to more complicated nuclear systems - provided the wave functions are known.
For the moment we   provide some results  for the p-shell nuclei for illustrative purposes.  
The information required for the six-quark clusters is the following:
The  number of radial quantum numbers:
\barr
n_1n_2&=&\{(0, 0), (0, 1), (0, 2), (1, 1), (0, 3), (1, 2), (0, 4), (1, 3), (2,2), (0, 5),\nonumber\\
&&  (1, 4), (2, 3)\} 
\earr
The radial overlaps in the above basis
\barr
C_{0s0s}^{n_1,n_2}&=&\{0.245178,0.468387,0.408421,0.223701,0.344056,0.39012,\nonumber\\
& &\hspace{-.5cm}.284612,0.328642,0.170088,0.232808,0.271861,0.28656 \}
\earr
\barr
C_{0s1s}^{n_1,n_2}&=&\{-0.234193,-0.127658,-0.0275665,-
0.121938,0.0473272,\nonumber\\&&\hspace{-1.5cm}-0.026332,0.097511,0.045207,-0.022960,0.01275,-0.074095,0.039419\}\nonumber
\earr
The coefficients:
 $ A_{0s}(01)=0.990\frac{1}{3}=0.331$, $ A_{1s}(01)=0.035\frac{1}{2}\frac{1}{3}=0.0331$, \\
 $ A_{0s}(10)=0.94\left ( -\frac{\sqrt{5}}{3} \right )=-0.78582$, $ A_{1s}(10) = 0.035 \frac{1}{2}\frac{1}{3}=0.00583333$\\
 (see section \ref{sec:snug}.)\\
 Twelve  nuclear matrix elements are required in  the coupling of the six-quarks to the above  nuclei, They were obtained using  the one-pion exchange potential given in ref  \cite{LakPand81}. They are as follows:\\
 \barr
 \mbox{ME(He)}&=&\{1.58831,-0.14934,-0.309879,0.442238,-0.254539,0.697035,\nonumber\\& & \hspace{-1.2cm}0.0669078,-0.0147065,0.230136,0.115068,0.0304126,-0.00588259\}\nonumber
 \earr
  \barr
 \mbox{ME(Li)}&=&\{1.61472,-0.151824,-0.315032,0.449593,
 -0.258772,0.708627,\nonumber\\
 &&\hspace{-1.5cm}0.0680205,-0.014951,0.233963,
 0.116982,0.0309184,-0.00598042\}
 \earr
 The sums in Eqs \ref{Eq:sum01} and \ref{Eq:sum10} are Num(He)= 0.184011 and Num(Li)=-0.452177, respectively.\\
 From nuclear mass tables  we find that BE($^4${He})=28.296, BE($^6$Li) = 31.995 and our calculations indicate E($q^6$(0,1)-2$E_{3q}(1/2.1/2)$=-7.95 MeV  and E($q^6$(1,0)-2 $E_{3q}(1/2,1/2)$=-4.71MeV. On the other hand $\beta$-decay experiments \cite{Knecht12} indicate that the energy of $^6$He is 5.5 MeV higher than that of $^6$Li. Based on that information we find 
 $D_{0,1}=9.8$ MeV and $D(1,0)=4.3$ MeV which results in $C_{0,1}=\frac{0.1811}{9.8}=0.019$ and $C_{1,0}=\frac{0.452}{4.3}=0.106$. Thus the probability of finding the six-quark cluster is:\\
 \beq
 P(\mbox{He})=3.5 \times 10^{-4},\quad P(\mbox{Li})=1.1 \times 10^{-2}.
 \eeq

Before embarking on more detailed calculations, however, ideally we would have  a good experimental signature to enable us detect the presence of the six-quark clusters in nuclei. The most appropriate  is the break up of the above nuclei into the closed-shell nucleus ($^4$He or $^{16}$O) and a six-quark cluster to unravel the exotic structure given, {\it e.g,}  in tables \ref{tab:exotic01} and \ref{tab:exotic10}.
For the moment we concentrate on double-beta and the possible effects of nuclear admixtures of six-quark clusters in nuclear transitions  caused by very short range operators, not hindered by nuclear short-range correlations. \\

\section {The Role of Six-Quark Admixtures  in the  the $0\nu\beta\beta$  Mediated by Heavy Majorana Neutrinos}
\label{sec:Ccfpbeta}

We consider  two states which can undergo neutrinoless double-beta decay:
\beq
(A,Z)\rightarrow (A,Z+2)+e^-+e^-
\eeq
proceeding via the exchange of very heavy Majorana neutrinos involving the states.
Consider the nuclear transition:
\beq
| A J_i T_z \rangle \rightarrow| A J_f (T_z-2) \rangle
\eeq
where we use the nuclear convention of isospin $T_z|n\rangle=\frac{1}{2}|n\rangle,\,T_z|p\rangle=-\frac{1}{2}|p\rangle$. For simplicity we do not  exhibit  the isospin  $T$ the  states, only  their $T_z$.\\  

We will also consider two additional nuclear states 
$(A-2)J'_i(T_z-1)$, $ (A-2)J'_f(T_z-1)$ which via the six-quark cluster can couple to the $A$ states. We begin with   mixing to the initial state. {\it i.e.} 
$$|i\rangle=|A J_i T_z\rangle +C_{J_i,J'_i}|(A-2)J'_i(T_z-1)\times q^6 0,1,1\rangle.$$
 The mixing coefficient is given by:
\barr
C_{J_i,J'_i}&=&\frac{1}{D_i}\langle (A-2)J'_i(T_z-1) \times q^6 0,1,1| V|  A J_i T_z\rangle,\nonumber\\D_i&=&E(A J_i T_z)-E((A-2)J'_i(T_z-1))-(E_B-2E_N)
\earr
In the energy denominator, in addition to the binding nuclear energies, the energy of the six-quark state relative to the two nucleons also appears.

Furthermore
\barr
& &\langle  (A-2)J'_i(T_z-1)\times q^6 0,1,1 |V| A J_i T_z\rangle\nonumber\\
&=&\sum_j\left ( \langle (A-2)J'_i(T_z-1);j^2J=0| \}A J_i T_z\rangle \right )\delta_{J,J'}\nonumber\\&&\langle  q^6 0,1,1|V|j^2J=0\rangle\nonumber
\earr 
where the coefficient $\langle (A-2)J'_i(T_z-1);j^2J=0| \}A J_i T_z\rangle$ is  the usual  coefficient of fractional parentage (cfp) in symbolic form where where $| j^2J=0 >$ is a short hand notation of the shell model states $| [(n \ell j)(n' \ell' j)]J=0> $.

Using the results of section \ref{sec:snug}, we find (omitting the $\ell=0$ in the bra)
\barr
&&\langle \Phi(6q)J_c=0,I_c=1,(I_c)_3=1|V|j^2J=0\rangle=\nonumber\\
\sum_{n_1,n_2,n \ell,n'\ell',j} &&\hspace{-.8cm}\Lambda_{n_1,n_2}(0,1)\langle (n_1\frac{1}{2}) (n_2 \frac{1}{2}) J=0,I=1|V|(n \ell j)(n' \ell' j),J=0 T=1\rangle\nonumber
\earr 

\barr
C_{J_i}&=&\frac{1}{D_i}\sum_j \langle (A-2)J_i(T_z-1);[(n \ell j)(n' \ell' j)]J=0| \}A J_i T_z\rangle\nonumber\\
&&\sum_{n_1,n_2,n \ell,n'\ell',j} \Lambda_{n_1,n_2}(0,1)\langle [(n_\frac{1}{2}) (n_2\frac{1}{2})] J=0,I=1|V|[(n \ell j)(n' \ell' j)]J=0 T=1\rangle\nonumber
\label{Eq:Mixingi}
\earr
Proceeding in an analogous fashion we can find the admixture to the final state:
$$|f\rangle=A J_f (T_z-2)+C_{J_f,J'_f}(A-2)J'_f(T_z-1)\times q^6 0,1,-1 >.$$ The mixing coefficient is given by:
\barr
C_{J_f}=\frac{1}{D_f}\sum_j &&\langle (A-2)J_i(T_z-1);[(n \ell j)(n' \ell' j)]J=0| \}(A J_f (T_z-2))\rangle\nonumber\\
\times  \hspace{-.3cm}\sum_{n_1,n_2,n \ell,n'\ell',j}\hspace{.2cm}&& \hspace{-1cm}\Lambda_{n_1,n_2}(0,1)\langle [(n_1\frac{1}{2}) (n_2\frac{1}{2})] J=0,I=1|V|[(n \ell j)(n' \ell' j)]J=0 T=1\rangle\nonumber\\
\label{Eq:Mixingf}
\earr
with
\barr
D_i&=&E(A J_i T_z)-E((A-2)J_i(T_z-1))-(E_B-2E_N),\nonumber\\D_f&=&E(A J_f (T_z-2))-E((A-2)J_f(T_z-1))-(E_B-2E_N)
\earr

\subsection {Evaluation of the $0\nu\beta\beta$  Matrix Elements in the Six-Quark Cluster}

\begin{figure}[h!]
	\begin{center}
		\includegraphics[width=0.45\textwidth]{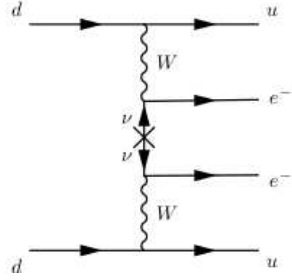}
		\caption{\footnotesize
			The elementary diagram at the quark level leading to neutrinoless double-beta decay in the case of heavy Majorana neutrino mechanism. Two down quarks with momenta $q_1$ and $q_2$ are transform into two up quarks with momenta $q'_1$ and $q'_2$ with the simultaneous emission of two electrons with momenta $p_1$ and $p_2$.} 
		\label{Fig:betabetaqq}
	\end{center}
\end{figure}

The  $0\nu\beta\beta$ transition can now be computed once the  nuclear wave functions are known. Here we will focus on the contribution coming  from the six-quark contribution. The nuclear matrix element involved will be of the form:
%\barr
\beq
|\mbox{ME}|^2=\left (C_{J_f}C_{J_i}  \right)^2 \;\left |\;\langle q^6 (00)_c1,1,-1|\Omega| q^6 (0 0)_c1,1,1\rangle \;
\right |^2
\eeq
%\earr
where $\Omega$ is the transition operator at the quark level resulting from the diagram in Fig.\ref{Fig:betabetaqq}.
The amplitude ${\cal M}'$ associated with this diagram in momentum space is inversely proportional to the neutrino mass $m_N$. Transforming this into coordinate space using the momentum conservation $\delta$-function $\delta({\bf q_1 + q_2-q'_1- q'_2-p_1-p_2})$ we find 
\beq
{\cal M}=\frac{1}{m_N}\delta \left (\bf{r}_1-\bf{r}_2\right ) e^{i(\bf{p}_1+\bf{p}_2). \bf{r}_2} =\frac{1}{b^3_N m_N}\omega,\,\omega=b^3_N\delta \left (\bf{r}_1-\bf{r}_2\right ) e^{i(\bf{p}_1+\bf{p}_2). \bf{r}_2}
\eeq
where $b_q$ is the scale parameter of the six-quark cluster introduced to make  $\omega$ dimensionless (see Eq. (\ref{Eq:OmegaFGT}) for the spin-isospin dependence  of $\omega$, namely $\Omega_F$ and  $\Omega_{GT}$). 

The normalization is consistent  with the radial factor one finds in the case of light neutrinos:
\beq
\frac{1}{M_N}\delta \left (\bf{r}_1-\bf{r}_2\right )\leftrightarrow\frac{M_N}{ 4 \pi r}e^{-m_N r} .
\eeq

Absorbing the factor $\frac{1}{b^3_N m_N}$ and the leptonic matrix element into the expression for the decay width we can write the transition ME as 
\barr
|\mbox{ME}|^2&=&\left (C_{J_f}C_{J_i}  \right)^2\left (\left |\langle  q^6 1,1,-1|\Omega_F|
 q^6 1,1,1\rangle 
\right |^2 \right .\nonumber\\
&+&\left .\left |\langle  q^6 1,1,-1|\Omega_{GT}|
 q^6 1,1,1\rangle 
\right |^2\right )
\earr
where in the long-wavelength approximation:
\beq
\Omega_F=\sum_{i\ne j}b^3_N\delta \left (\bf{r}_i-\bf{r}_j\right 
)\tau_-(i) \tau_-(j),\, \Omega_{GT}=\sum_{i\ne j}b^3_N\delta 
\left (\bf{r}_i-\bf{r}_j\right )\tau_-(i) \tau_-(j)\sbf_i.\sbf_j
\label{Eq:OmegaFGT}
\eeq
(at the quark level $g_A=1$). The required matrix elements can  be obtained in the context of the six-quark cluster.

Let us consider the leading $0s^6$ configuration. Using the results of ref. \cite{dds2}, we have

$$  \langle  q^6 1,1,-1|\Omega| q^6 1,1,1\rangle =\frac{1}{2}\times 6\times 5$$
$$ \langle 0,0 ;1,1|1,1\rangle^2 \left( \frac{1}{4} ME_0+\frac{1}{12} ME_1 \right ) + \langle 2,0 ;1,1|1,1\rangle^2 \left (\frac{1}{10} ME_0+\frac{1}{15}ME_1)  \right ) .$$ 

Due to isospin conservation the two-body ME can only involve $I=1$, {\it i.e.} only the $[1^2]$ color-spin symmetry is relevant. The first factors outside the parenthesis are the relevant isospin Clebsch-Gordan coefficients taking the values of 1 and $\frac{1}{10}$. In the above expression 
$$ME_s=\langle 0s^2 I=1,I_3=-1,s|\omega|0s^2 I=1,I_3=1,s\rangle$$
In the case of Fermi interaction $ ME_0=ME_1=R_{0s}$. while for the Gamow-Teller interaction $ME_0=-3 R_{0s}$ and $ME_1=R_{0s}$. Due to the presence of the $\delta$ function in the interaction the radial integral becomes
\beq
R_{0s}=\frac{1}{4 \pi}\int_0^\infty 0s^4(x) x^2 dx=\frac{1}{( 2 \pi)^{3/2}}
\eeq

We thus obtain
\barr
\langle  q^6 1,1,-1|\Omega_F| q^6 1,1,1\rangle &=&\frac{21}{4}\frac{1}{( 2 \pi)^{3/2}},\nonumber\\
\langle  q^6 1,1,-1|\Omega_{GT}| q^6 1,1,1\rangle &=&-\frac{207}{20}\frac{1}{( 2 \pi)^{3/2}}\nonumber\\
\earr
Thus total matrix element due to the $0s^6$  contribution only for both the Fermi and Gamow-Tellar transitions becomes:
\beq
\left (\mbox{ME}^{\beta\beta}_{0s^6}\right )^2=\frac{26937}{1600 \pi^3}\approx 0.54
\eeq

Using this nuclear ME the decay width can be obtained with the standard nuclear kinematics and it can be cast in  the form:
\barr
&&d \Gamma=\left (\frac{G_F}{\sqrt{2}}\right )^4|\mbox{ME}|^2\nonumber\\&&\left(\frac{1}{b^3_N  
	m_N} \right)^2  \left ( 4\frac{E_1 E_2-{\bf p_1.p_2}}{m_e^2}\right )\frac{1}{(2 
	\pi)^5}
d^{3}{\bf p}_1  d^{3}{\bf p}_2 \delta \left (\Delta-E_1-E_2  
\right )
\earr
where $\Delta$ is the available energy and $E_1$ and $ E_2$ the 
total energies of the outgoing electrons.

Since the phase space part is the same with the standard mechanism, we can write the six-quark neutrinoless  
double-beta decay life time associated with the six-quark cluster  as:
\beq
\left [ T_{1/2}(0\nu)\right ]^{-1}=\eta^2_{N}\left(ME(6q) 
\right)^2_{eff} G_{0\nu}(E_0,Z)
\eeq
where
\beq
\eta_{N}=\sum_j U^2_{ej}\frac{m_p}{m_{N_j}}
\eeq
is  the lepton violating parameter as usual,
\barr
\left(ME(6q)\right)_{eff}^2&=&\left (C_{J_f}C_{J_i}  
\right)^2 f_{sc}\nonumber\\&&\left (\left |\langle q^6 1,1,-1|\Omega_F|
 q^6 1,1, 1\rangle 
\right |^2 \right .\nonumber\\
&+&\left .\left |\langle  q^6 1,1,-1|\Omega_{GT}|
 q^6 1,1,1\rangle 
\right |^2\right )\nonumber\\
\earr
is the effective nuclear transition matrix element arising from the six-quark cluster and $ G_{0\nu}(E_0,Z)$ is the phase space factor. The scale factor ) $f_{sc}$ is necessary since the phase space factor is normally given in the context of light neutrino mass mechanism. It is absorbed into the effective nuclear matrix element.  In the case of a heavy neutrino it is given by
\beq
\left(\frac{1}{b^2_q  
	m_e m_p}\frac{R_0}{b_q} \right)^2\approx 3  \times 10^5 A^{2/3}
\eeq
\subsection{Evaluation of the Double-Beta Decay Matrix Element for  $A=48$}

We  consider for simplicity the A=48 system within the $f_{7/2}$ shell assuming an initial state,
 $0f_{7/2}^8$ neutrons with $J_i=0$.\\
The allowed mass 46 $ |J'_i,T'_i>$ states  are $J'_i=0,2,4,6,\,T'_i=1$ with the cfp's in the spirit of  Eq. (\ref{Eq:Mixingi}) being $C_{J'}= \frac{\sqrt{2 J_i'+1}}{2\sqrt{7}}$. Only the $J'=0$ is relevant here with a  cfp$= 1/(2\sqrt{7})$, which unfortunately is quite small.\\
%$\sqrt{1/15},\sqrt{1/3}$ and $\sqrt{3/5},$ respectively. \\
The final state is of very simple form,  a product  0f $0f_{7/2}^6J'_f$ neutrons and $0f_{7/2}^2 J'_f$ protons coupled to a total angular momentum $J_f=0$. The corresponding amplitudes for $J'_f =0,2,4,6$ can be obtained by a simple shell model calculation, {\it e.g.} $ 0.845,-0.523,0.109,-0.001,$ respectively found in ref. \cite{VerBB83}. The relevant amplitude is for $J'_f =0$ or 0.845.

From nuclear mass tables we find the binding energies are for $^{48}$Ca $ ^{46}$Ca and $^{48}$Ti  $415.991,\,398.769, and\,420.191$ MeV, respectively.  
% our calculation shows that for the $J=0,I=1$ 6q cluster $E_b-2E_N=-6$ MeV
 Taking the quark  energy difference  the same as for $^6$He, that is E($q^6$(0,1)-2$E_{3q}$(1/2,1/2)=-7.95 MeV, the energy denominators become 
\beq
D_i= -9.27\;\mbox{MeV},\,D_f=-16.71\;\mbox{MeV .}
\eeq
Using the coefficients $ A_{0s}(01)=0.990\frac{1}{3}=0.331$, $ A_{1s}(01)=0.035\frac{1}{2}\frac{1}{3}=0.0331$, see section \ref{sec:snug}. Proceeding as in section \ref{sec:MicBSM} we find the  twelve needed nuclear matrix elements
$$\mbox{ME}(^{48}\mbox{Ca})= \{0.16938,0.103112,0.193568,0.281692,0.00770124,-0.00769442,$$ $$-0.04825,-0.0610827,-0.0225847,-0.024125,-0.0277649,-0.00903389\}$$
which are indeed very small.
 Hence we find for the sum appearing in  the numerator of Eq. (\ref{Eq:Mixingi}) or (\ref{Eq:Mixingf}) the value of 0.0584 MeV. Thus
\beq
C_i = \frac{0.0584}{D_i} \frac{1}{2 \sqrt{7}}=-0.00119061,\,C_f = \frac{0.0584}{D_f} 0.845=-0.00295366\;.
\eeq
Thus in the approximation of a purely $0f_{\frac{7}{2}}$ configuration, the probability of finding  the $J=0,I=1$ six-quark cluster in the $^{48}$Ca nucleus is extremely small
\beq
C_i^2=1.4\times 1)^{-6} \mbox{ for }^{48}\mbox{Ca},\,C_f^2=8.7\times 10^{-6} \mbox{ for }^{48}\mbox{Ti}.
\eeq
The resulting nuclear matrix element is small
\beq
ME_{0\nu \beta \beta}=C_iC_f \sqrt{0.54}f_r=0.0047
\eeq
The factor $f_r=m_N/m_e=1.8 \times 10^3$ was employed since traditionally, this is how the   $0\nu\beta\beta$ matrix elements for heavy neutrino exchange are normalized. This is a tiny  value compared with the standard phenomenological treatment\cite{Vergados82}:
\beq
ME_{0\nu \beta \beta}=72
\eeq
\clearpage
\section{Concluding Remarks}
\label{sec:ConRem}
In this paper we examined the structure of six-quark clusters in the context of the group $SU(6)_L\otimes SU(6)_{cs}\otimes SU(2)_I$ the first corresponding to orbital, the second to color-spin and the third to isospin symmetry. We considered a model involving harmonic oscillator single quark states within two $\hbar \omega$ with all color singlet states and with all allowed spin and isospin states, focusing on those of the two  nucleons, {\it i.e.} color singlet  states$J=0,I=1$ and $J=1,I=0$. The Hamiltonian matrix was diagonalized using the interaction derived from the one-gluon exchange potential and various other phenomenological quark-quark interactions. The wave functions were expanded into two $q^3$ clusters. To this end we computed all the one-particle, two-particle and  three-particle cfp's for all the three symmetries involved.

The probability of two-nucleon components in mass 48  was about 1/9 and 5/8 for the $J=0,I=1$ and $J=1,I=0$, respectively. The remainder corresponds to other baryon-like states or colored three-quark components. Using  only the two nucleon-like component, we obtained the coupling of the six-quarks to  two-nucleon shell model states and thus the probability of finding the six-quark structure in a nucleus. This was straightforward for the nuclei that are considered as two particles outside a closed shell. For such simple cases this probability is relatively large, of about $1 \%$ in the case of $^{6}$Li.
As the mass number increases the spin-orbit interaction becomes more important and this tends to destroy LS coupling.  But isolated areas of the mass table can be found in which the low $\ell$ and $j$ can dominate the single-particle wave functions.\\ 

For more complicated nuclei it may be less suppressed due to the $\langle A-2|\}A\rangle$ nuclear coefficients of fractional parentage involved.  To test for a signature of the presence of six-quark cluster in nuclei we considered a process involving a very short-range transition  operator, in particular the neutrinoless double-beta decay mediated by very heavy Majorana neutrinos.  We found that indeed this proceeds even with a delta-function interaction between the quarks inside the six-quark bag. As a test we considered the simple double-beta decay  $^{48}$Ca to $^{48}$Ti. The resulting neutrinoless double-beta decay nuclear matrix element was found, however,  to be very small, mainly  due to the smallness of the nuclear coefficients of fractional parentage involved.
%where the first number corresponds to nuclear binding energy differences and the second to the estimated cluster energy.

%%%%%%%%%%%%%%%%%%%%%%%%%%%%%%%%%%%%%%%%%%%%%%%%%%%%%%%%%%%

%\bibliographystyle{ws-rv-van}
%\bibliography{bibfile}
%\bibliographystyle{ws-ijmpe}
%\bibliography{Bib}

\clearpage

\section{Radial Overlaps}

One must now expand the radial wave functions from $(\hbar \omega)_q$  to the $(\hbar \omega)_N$ employed in the nuclear shell model.
Thus,
\beq
\big | 0s({ \bf r}_1)0s({ \bf r}_2) > =\sum_{n1\le n_2} C^{0s,0s}_{n_1n_2} \;\big | (n_1,0)(n_2,0)L=0\rangle,\,C^{0s,0s}_{n_1n_2}=\langle n_1,0|0s\rangle\langle n_2,0|0s\rangle .\nonumber
\eeq
with similar equations for the other two-quark states.  
The expansion does not converge very fast, but only a few terms in the sum are adequate for our purposes.  Many radial components appear:
\barr
|f_{1,0}\rangle&=&\sum_{n1,n2} \left (A_1|C^{0s0s}_{n_1,n_2}+A_2|C^{0s1s}_{n_1,n_2}\right )\nonumber\\ 
&&\times |(n_1,\ell_1=0),(n_2,\ell_2=0) ,L=0,S=0,J=0,I=1 \rangle 
\label{Eq:expn1n2}
\earr

\begin{figure}[htbp]
	\center
	\includegraphics[width=0.45\textwidth]{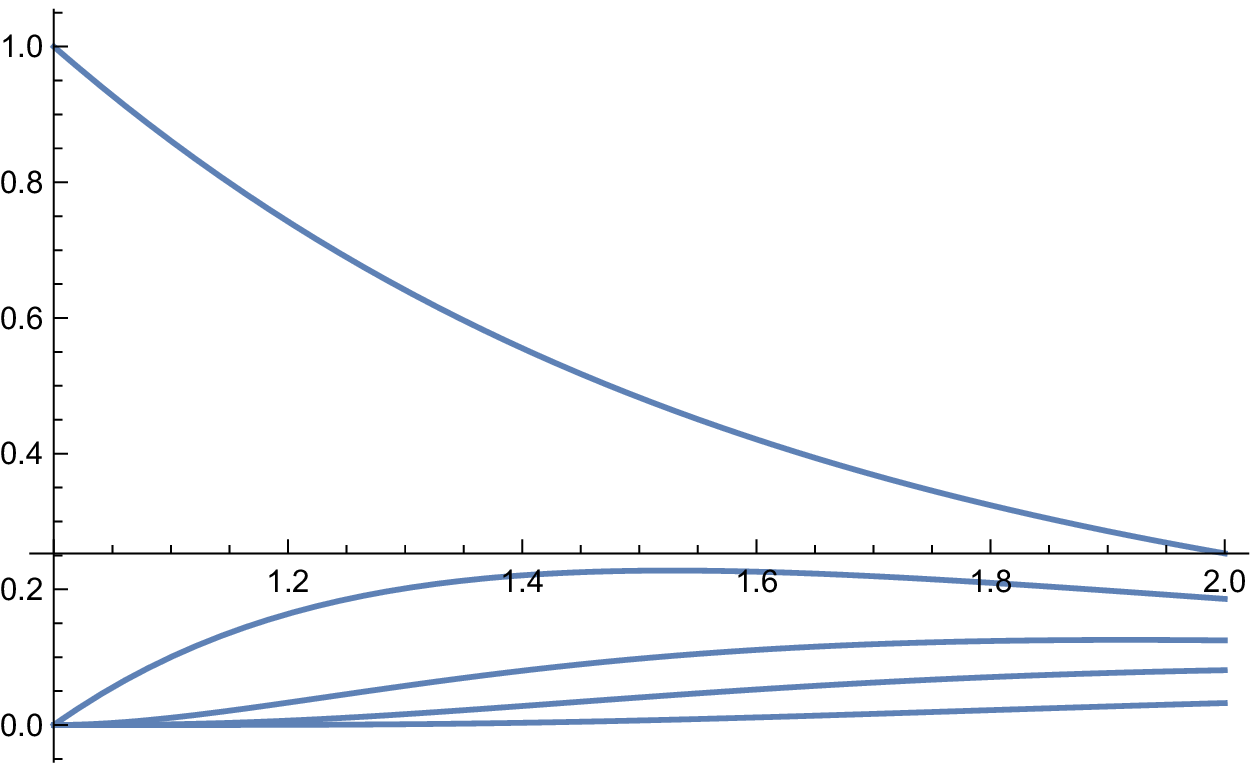}
	\includegraphics[width=0.45\textwidth]{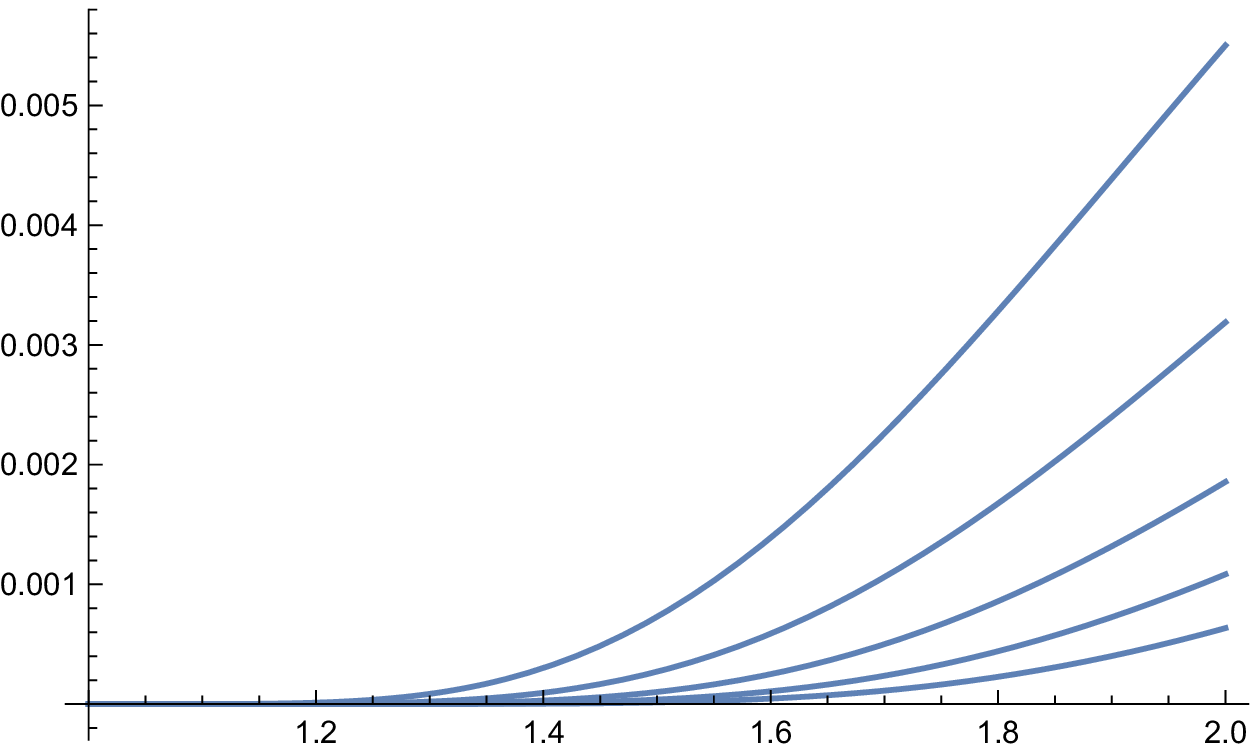}
	\caption{The overlap integrals $\langle 0s|n_i,\ell=0\rangle, i=0,1,\cdots 4 \mbox{(left)},i=5,6,\cdots10\mbox{ (right)}$, as a function of $b_A/b_q$, with $b_a$ being the harmonic oscillator length parameter for the nucleon and $b_q$ that of the quarks in the six-quark cluster.
		In the curves the larger the $n_i$ the smaller the value of the overlap.}
	\label{fig:radovlp}
\end{figure}
Below we list these coefficients up to $ n_1+n_2=5$, an approximation that appears reasonable. If larger values of $  n_1+n_2$ are needed one can use the tables of Appendix D in section \ref{AppendixRaD}
\begin{table} [ht]
	\caption{$C^{0s0s}_{n_1,n_2} $and $C^{0s1s}_{n_1,n_2}$ coefficients.
		\label{tab:n1n2J0}}
	$\frac{b_A}{b_q}=1.5$\\
	$$
	\left(
	\begin{array}{ccc}
	\{n_1,&n_2\}&C^{0s0s}_{n_1,n_2}\\
	0 & 0 & 0.786527 \\
	0 & 1 & 0.370498 \\
	0 & 2 & 0.159319 \\
	1 & 1 & 0.174525 \\
	0 & 3 & 0.066186 \\
	1 & 2 & 0.075048 \\
	0 & 4 & 0.0270004 \\
	1 & 3 & 0.0311774 \\
	2 & 2 & 0.0322717 \\
	0 & 5 & 0.0108917\\
	1 & 4 & 0.0127187 \\
	2 & 3 & 0.0134067 \\
	\end{array}
	\right),\,
	\left(
	\begin{array}{ccc}
	\{n_1,&n_2\}&C^{0s1s}_{n_1,n_2}\\
	0 & 0 & -0.370498 \\
	0 & 1 & 0.495652 \\
	0 & 2 & 0.501322 \\
	1 & 1 & 0.23348 \\
	0 & 3 & 0.327986 \\
	1 & 2 & 0.23615 \\
	0 & 4 & 0.182641 \\
	1 & 3 & 0.154500 \\
	2 & 2 & 0.101548 \\
	0 & 5 & 0.0334612 \\
	1 & 4 & 0.0165542 \\
	2 & 3 & 0.0293474 \\
	
	\end{array}
	\right)
	$$
	$ \frac{b_A}{b_q}=2.0$\\
	$$
	\left(
	\begin{array}{ccc}
	\{n_1,&n_2\}&C^{0s0s}_{n_1,n_2}\\
	
	0 & 0 & 0.51200 \\
	0 & 1 & 0.752483 \\
	0 & 2 & 0.504781 \\
	1 & 1 & 0.27648 \\
	0 & 3 & 0.327136 \\
	1. & 2 & 0.370937 \\
	0 & 4 & 0.208188 \\
	1 & 3 & 0.240395 \\
	2 & 2 & 0.124416 \\
	0 & 5 & 0.13101 \\
	1 & 4 & 0.152986 \\
	2 & 3 & 0.161262 \\
	\end{array}
	\right)
	\,
	\left(
	\begin{array}{ccc}
	\{n_1,&n_2\}&C^{0s1s}_{n_1,n_2}\\
	0 & 0 & -0.376242 \\
	0 & 1 & 0.051200 \\
	0 & 2 & 0.25416 \\
	1 & 1 & 0.0376242 \\
	0 & 3 & 0.307171 \\
	1 & 2 & 0.186769 \\
	0 & 4 & 0.286141 \\
	1. & 3. & 0.225724 \\
	2 & 2 & 0.125288 \\
	0 & 5 & 0.237114 \\
	1 & 4 & 0.0104094 \\
	2 & 3 & 0.1514200 \\
	\end{array}
	\right)
	$$
	$ \frac{b_A}{b_q}=2.8$\\
	$$
	\left(
	\begin{array}{ccc}
	\{n_1,&n_2\}&C^{0s0s}_{n_1,n_2}\\
	0 & 0 & 0.245178 \\
	0 & 1 & 0.468387 \\
	0 & 2 & 0.408421 \\
	1 & 1 & 0.223701 \\
	0 & 3 & 0.344056 \\
	1 & 2 & 0.390123 \\
	0 & 4 & 0.284612 \\
	1 & 3 & 0.328642 \\
	2 & 2 & 0.170088 \\
	0 & 5 & 0.232808 \\
	1 & 4 & 0.271861 \\
	2 & 3 & 0.286567 \\
	\end{array}
	\right)
	\,
	\left(
	\begin{array}{ccc}
	\{n_1,&n_2\}&C^{0s1s}_{n_1,n_2}\\
	0 & 0 & -0.234193 \\
	0 & 1 & -0.127658 \\
	0 & 2 & -0.0275665 \\
	1 & 1 & -0.121938 \\
	0 & 3 & 0.0473272 \\
	1 & 2 & -0.0263315 \\
	0 & 4 & 0.0975105 \\
	1 & 3 & 0.0452068 \\
	2 & 2 & -0.0229604 \\
	0 & 5 & 0.1275000 \\
	1 & 4 & -0.074095 \\
	2 & 3 & 0.0394191 \\
	\end{array}
	\right)
	$$
\end{table}
\clearpage

\section{Original $\kappa$ tables}

\begin{table}[h]\centering
	\begin{tabular}{|c|c|c|c|c|c|c|c|c|c|c|c||} 
		\hline
	$0s^6$ & $[f]_{cs}$ & $(\lambda,\mu)$ &&&&&& &$S$ & $L$ & $\kappa^{0s^6}_{NN }$  \\\hline
		1(1) & $[2^21^2]$ & $(0,0)$ &&& &&&& 0 & 0 & $1/2 $\\		
		\hline\hline
		$0s^51s $ & $[f]_{cs} $& $(\lambda_1,\mu_1) $ & $ S_1$& $I_1$&$[f_2]$ & $(\lambda_2,\mu_2$ & $S_2$ & $I_2$ & $S $& $L$ & $\kappa^{0s^51s}_{NN} $ \\\hline
	  $	2(1)$ & $[2^21]$ & $(0,1)$ & $1/2$ &1/2& $[1]$ & $(1,0)$ & $1/2 $ &1/2 & 0 &  0& $5/9$\\
	$3(2)$ & $[21^3]$ & $(0,1)$ & $1/2$ &1/2& $[1]$ & $(1,0)$ & $1/2$ & 1/2 & 0 & 0 & $-4/9$\\							\hline\hline				
		$0s^50d$ & $[f]_{cs}$ & $(\lambda_1,\mu_1)$ & $S_1$ & $I_1$ & $[f_2]$ & $(\lambda_2,\mu_2$ & $S_2$ & $I_2$ &$S$ & $L$ & $\kappa^{0s^50d}_{NN}$\\\hline
		4(1)  & $[2^21] $ &(0,1)&3/2& 1/2&[1]&(1,0)&1/2&1/2&2&2&0\\
		5(2)  & $[2^21] $ &(0,1)&5/2& 1/2&[1]&(1,0)&1/2&1/2&2&2&0\\
		6(3)  & $[21^3] $ &(0,1)&3/2 & 1/2&[1]&(1,0)&1/2&1/2&2&2&0\\
		7(4)  & $[21^3] $ &(0,1)&5/2& 1/2&[1]&(1,0)&1/2&1/2&2&2&0\\ \hline				
%	\end{tabular}
%	\begin{tabular}{|c|c|c|c|c|c|c|c|c|c|c|lc|c||c|c||}
		\hline
		$0s^4 0p^2$ & $[f]_{cs} $&$ (\lambda_1,\mu_1)$ & $S_1$ & $I_1$ & $[ f ]_2$ & $(\lambda_2,\mu_2)$ & $S_2$ & $I_2$ & $S$ & $L$ & $\kappa^{0s^40p^2}_{NN}$  \\\hline
		8(1) & $[2^2]$ &(0,2) &0 &0& $[1^2]$ &(2 0)&0&1&0&0&0  \\
		9 (2) &$[2^2]$& (1,0) &1&0& $[1^2]$ &(0,1)&1&1&0&0&$\sqrt{6}/9$  \\
		10	(3)&$[2^2]$&(0,2)&2&0& $[1^2]$ &(2,0)&0&1&2&2&0  \\
		11(4)& $[2^2]$ &(1,0)&1&0& $[1^2]$ &(0,1)&1&1&2&2&0  \\
		12(5)& $[21^2]$     &(1,0)&1&1& $[1^2]$ &(0,1)&1&1&0&0&1/3  \\
		13(6) &$[21^2]$ &(0,2)&1&1&[2]&(2,0)&1&0&0&0&0  \\
		14	(7)& $[21^2]$   &(1,0)&0&1&[2]&(0,1)&0&0&0&0&$\sqrt{3}/6$  \\
		15	(8)& $[21^2]$ &(1,0)&1&1& $[1^2]$ &(0,1)&1&1&2&2&0  \\
		16(9)& $[21^2]$ &(1,0)&2&1& $[1^2]$ &(0,1)&1&1&2&2&0  \\
		17	(10)& $[21^2]$ &(0,2)&1&1&[2]&(2,0)&1&0&2&2&0  \\
		18(11)& $[21^2]$ &(1,0)&2&1&[2]&(0,1)&0&0&2&2&0  \\
		19(12) &$[1^4$]&(0,2)&0&2& $[1^2]$ &(2,0)&0&1&0&0&0  \\
		20(13) &$[1^4]$&(1,0)&1&2& $[1^2]$ &(0,1)&1&1&0&0&0  \\
		21(14) &$[1^4$]&(1,0)&1&2& $[1^2]$ &(0,1)&1&1&2&2&0  \\
		\hline
		22	(1)&$[2^2]$&(0,2)&0&0&[2]&(2,0)&1&1&1&1&0  \\
		23			(2)&$[2^2]$&(0,2)&2&0&[2]&(2,0)&1&1&1&1&0  \\
		24(3)&$[2^2]$&(0,2)&2&0&[2]&(2,0)&1&1&2&1&0  \\
		25				(4)&$[2^2]$&(1,0)&1&0&[2]&(0,1)&0&1&1&1&1/3  \\
		26(5)&$[2,1^2]$ &(0,2)&0&1&[2]&(2,0)&1&1&1&1&0  \\
		27	(6)&$[2,1^2$ ]&(0,2)&1&1&[2]&(2,0)&1&1&1&1&0  \\
		18	(7)&$[2,1^2]$ &(1,0)&1&1&[2]&(0,1)&0&1&1&1&-$\sqrt{3}/3$  \\
		29	(8)&$[2,1^2]$ &(0,2)&1&1& $[1^2]$ &(2,0)&0&1&1&0 & \\
		30	(9)&$[2,1^2]$ &(1,0)&0&1& $[1^2]$ &(0,1)&1&0&1&1&-$\sqrt{6}/6$  \\
		31	(10)&$[2,1^2]$ &(1,0)&1&1& $[1^2]$ &(0,1)&1&0&1&1&-$\sqrt{3}/3$  \\	
		\hline
	\end{tabular}
	\caption{The $I=1,\,J=0$  basis six-quark states and the spin-isospin  coefficients $\kappa^{0s^6}_{NN}$,  $\kappa^{0s^51s(0d)}_{NN}$ and $\kappa^{0s^40p^2}_{NN}$ needed to transform them into two nucleon-like configurations.  The orbital part is given in the text.}
	\label{tab:taba2}
\end{table}

% ii) {$J=1,I=0$ states}\\
\begin{table}[h]\centering
	\begin{tabular}{|c|c|c|c|c|c|c|c|c|c|c|c|}
		\hline
	$0s^6$ & $[f]_{cs}$ & $(\lambda,\mu)$ &&&&&&& $S$ & $L$ & $\kappa^{0s^6}_{NN }$  \\\hline
		1(1)&$[2^3]$&(0,0)&&&&&&&1&0&$\sqrt{5}/6$\\\hline\hline
		$0s^51s $ & $[f]_{cs} $& $(\lambda_1,\mu_1) $ & $ S_1$&$I_1$& $[f_2]$ & $(\lambda_2,\mu_2$ & $S_2$ & $I_2$& $S $& $L$ & $\kappa^{0s^51s}_{NN} $ \\\hline
		2(1)& $[2^21]$&(0,1)&1/2&1/2&[1]&(1,0)&1/2&1/2&1&0&5/9\\
		3(2)& $[2^21]$&(0,1)&3/2&1/2&[1]&(1,0)&1/2&1/2&1&0&$2\sqrt{5}/9$\\\hline
		$0s^50d $ & $[f]_{cs} $& $(\lambda_1,\mu_1) $ & $ S_1$&$I_1$&$[f_2]$ & $(\lambda_2,\mu_2$ & $S_2$ &$I_2$& $S $& $L$ & $\kappa^{0s^50d}_{NN} $ \\\hline
		4(1)& $[2^21]$ &(0,1)&1/2&11/2&[1]&(1,0)&1/2&1/2&1&2&5/9\\
		5(2)& $[2^21]$ &(0,1)&3/2&11/2&[1]&(1,0)&1/2&1/2&1&2&2$\sqrt{5}/9$\\
		6(3)& $[2^21]$ &(0,1)&3/2&11/2&[1]&(1,0)&1/2&1/2&2&2&0\\
		7(4)& $[2^21]$ &(0,1)&5/2&11/2&[1]&(1,0)&1/2&1/2&2&2&0\\
		8(5)& $[2^21]$ &(0,1)&5/2&1/2&[1]&(1,0)&1/2&1/2&3&2&0\\ \hline
%	\end{tabular}
%\end{table}
%
		\hline
		$0s^4 0p^2$ & $[f]_{cs} $&$ (\lambda_1,\mu_1)$ & $S_1$ & $I_1$ & $[ f ]_2$ & $(\lambda_2,\mu_2)$ & $S_2$ & $I_2$ & $S$ & $L$ & $\kappa^{0s^40p^2}_{NN}$  \\\hline
		9  (1)    & $[2^2]$ &(0,2)&0&0&[2]&(2,0)&1&0&1& 0&0  \\
		10      (2)   & $[2^2]$ &(0,2)&2&0&[2]&(2,0)&1&0&1&0&0  \\
		`11(3) & $[2^2]$ &(1,0)&1&0&[2]&(0,1)&0&0&1&0&-$\sqrt{2}/2$  \\
		12    (4)  & $[2^2]$ &(0,2)&0&0&[2]&(2,0)&1&0&1&2&0  \\
		13(5) & $[2^2]$ &(0,2)&2&0&[2]&(2,0)&1&0&1&2&0  \\
		14 (6)& $[2^2]$ &(0,2)&2&0&[2]&(2,0)&1&0&2&2&0  \\
		15(7) & $[2^2]$ &(0,2)&2&0&[2]&(2,0)&1&0&3&2&0  \\
		16(8) & $[2^2]$ &(1,0)&1&0&[2]&(0,1)&0&1&1&2&-$\sqrt{2}/2$  \\
		\hline	
		17(9) & $[2,1^2]$ &(0,2)&1&1& $[1^2]$ &(20)&0&1&1&0&0  \\
		18(10) & $[2,1^2]$ &(1,0)&1&1& $[1^2]$ &(0,1)&1&1&1&0&$\sqrt{6}/6$  \\
		19(11) & $[2,1^2]$ &(1,0)&1&1& $[1^2]$ &(0,1)&1&1&1&0&1/3  \\
		20(12) & $[2,1^2]$ &(1,0)&1&2& $[1^2]$ &(0,1)&1&1&1&0&0  \\
		21 (13) & $[2,1^2]$ &(0,2)&1&1& $[1^2]$ &(20)&0&1&1&2&0  \\
		22(14) & $[2,1^2]$ &(1,0)&0&1&0 $[1^2]$ &(0,1)&1&1&1&2&$\sqrt{6}/6$  \\
		23(15) & $[2,1^2]$ &(1,0)&1&1& $[1^2]$ &(0,1)&1&1&1&2&0  \\
		24(16)  & $[2,1^2]$ &(1,0)&1&1& $[1^2]$ &(0,1)&1&1&2&2&0  \\
		25(17)  & $[2,1^2]$ &(1,0)&2&1& $[1^2]$ &(0,1)&1&1&1&2&0  \\
		26 (18)    & $[2,1^2]$ &(1,0)&2&1& $[1^2]$ &(0,1)&1&1&2&2&0  \\
		27 (19)      & $[2,1^2]$ &(1,0)&2&1& $[1^2]$ &(0,1)&1&1&3&2&-$\sqrt{2}/2$  \\
		\hline
		28(20)&   $[2^2]$ 	&(0,2)&0&0& $[1^2]$ &(2,0)&0&0&0&1&0  \\
		29(21)&   $[2^2]$ 	&(0,2)&2&0& $[1^2]$ &(2,0)&0&0&2&1&0  \\
		30 (22)&   $[2^2]$ 	&(1,0)&1&0& $[1^2]$ &(0,1)&1&0&0&1&-$\sqrt{3}/3$  \\
		31 (23)  & $[2^2]$ 	&(1,0)&1&0& $[1^2]$ &(0,1)&0&0&1&1&-$\sqrt{2}/3$  \\
		32(24)&   $[2^2]$ 	&(1,0)&1&0& $[1^2]$ &(0,1)&1&0&2&1&0  \\
		33(25)&   $[2,1^2]$ 	&(0,2)&1&1&[2]&(2,0)&1&1&0&1&0  \\
		34(26)&   $[2,1^2]$ 	&(0,2)&1&1&[2]&(2,0)&1&1&1&1&0  \\
		35(27)&   $[2,1^2]$ 	&(0,2)&1&1&[2]&(2,0)&1&1&2&1&0  \\
		36(28)&   $[2,1^2]$ 	&(1,0)&0&1&[2]&(0,1)&0&1&0&1&-$\sqrt{2}/2$  \\
		37(29)&   $[2,1^2]$ 	&(1,0)&1&1&[2]&(0,1)&0&1&1&1&$\sqrt{6}/6$  \\
		38(30)&   $[2,1^2]$ 	&(1,0)&2&1&[2]&(0,1)&0&1&2&1&0  \\ \hline
	\end{tabular}
	\caption{The $I=0,\,J=1$  basis six-quark states and the spin-isospin coefficients $\kappa^{0s^6}_{NN}$,  $\kappa^{0s^51s(0d)}_{NN}$ and $\kappa^{0s^40p^2}_{NN}$ needed to transform them into two nucleon like configurations.The orbital part is given in the text.}
	\label{tab:tabb2}
\end{table} 

\section{Appendix A: Matrix elements of the interaction}
\label{sec.Appendix A}
Since the interaction depends on the relative coordinates it is useful to express the 2-quark shell model states in terms of relative and center of mass coordinates via the relations:
\beq
{\bf r}=\frac{1}{\sqrt{2}}({\bf r}_1-{\bf r}_2),\,{\bf R}=\frac{1}{\sqrt{2}}({\bf r}_1+{\bf r}_2)
\eeq
Then the states of interest in our work are:
$$|1\rangle=0s({\bf r}_1)0s({\bf r}_2)=0s({\bf r})0s({\bf R})$$
$$|2\rangle=\frac{1}{\sqrt{2}}\left (0s({\bf r}_1)1s({\bf r}_2)+0s({\bf r}_2)1s({\bf r}_1)\right )=\frac{1}{\sqrt{2}}\left (0s({\bf r})1s({\bf R})+0s({\bf R})1s({\bf r})\right )$$
$$|3\rangle=\left[ (0p({\bf r}_1)0p({\bf r}_2)\right ]^{L=0}=\frac{1}{\sqrt{2}}\left (0s({\bf r})1s({\bf R})-0s({\bf R})1s({\bf r})\right )$$
$$|4\rangle=\frac{1}{\sqrt{2}}\left (0s({\bf r}_1)0d({\bf r}_2)+0s({\bf r}_2)0d({\bf r}_1)\right )=\frac{1}{\sqrt{2}}\left (0s({\bf r})0d({\bf R})+0s({\bf R})0d({\bf r})\right )$$
$$|5\rangle=\left[ (0p({\bf r}_1)0p({\bf r}_2)\right ]^{L=2}=\frac{1}{\sqrt{2}}\left (0s({\bf r})0d({\bf R})-0s({\bf R})0d({\bf r})\right )$$
$$|6\rangle=\left[ (0p({\bf r}_1)0p({\bf r}_2)\right ]^{L=1}=\left[ (0p({\bf R})0p({\bf r})\right ]^{L=1}$$
$$|7\rangle=\frac{1}{\sqrt{2}}\left (0s({\bf r}_1)0p({\bf r}_2)+0s({\bf r}_2)0p({\bf r}_1)\right )=0s({\bf r})0p({\bf R})$$
$$|8\rangle=\frac{1}{\sqrt{2}}\left (0s({\bf r}_2)0p({\bf r}_1)-0s({\bf r}_1)0p({\bf r}_2)\right )=0s({\bf R})0p({\bf r})$$
$$|9\rangle=\frac{1}{\sqrt{2}}\left (1s({\bf r}_2)0s({\bf r}_1)-1s({\bf r}_1)0s({\bf r}_2)\right )=[0p({\bf R})0p({\bf r})]^{L=0}$$
(The 1s convention used by mathematica is positive at the origin, i.e $(3/2-x^2)$
$$|10\rangle=\frac{1}{\sqrt{2}}\left (0d({\bf r}_1)0s({\bf r}_2)-0d({\bf r}_2)0s({\bf r}_1)\right )=[0p({\bf R})0p({\bf r})]^{L=2}$$
Note that the states  $|9\rangle$ and $|10\rangle$ are antisymmetric, is spite of the fact that their orbital angular momentum is even.

The energy matrix is symmetric with non-zero ME being:
$$ \langle 1| \upsilon_s|1\rangle=\sqrt{2}\frac{1}{\sqrt{\pi}},\,\langle 1| \upsilon_s|2\rangle=-\langle 1| \upsilon_s|3\rangle=\frac{1}{\sqrt{6}}
\frac{1}{\sqrt{\pi}}$$
$$ \langle 2| \upsilon_s|2\rangle=\langle 3| \upsilon_s|3\rangle=\frac{11}{6 \sqrt{2}}\frac{1}{\sqrt{\pi}},\,\langle 2| \upsilon_s|3\rangle=\frac{1}{6 \sqrt{2}}
\frac{1}{\sqrt{\pi}}$$
$$ \langle 4| \upsilon_s|4\rangle=\langle 5| \upsilon_s|5\rangle=\frac{23}{15 \sqrt{2}}\frac{1}{\sqrt{\pi}},\,\langle 4| \upsilon_s|5\rangle=\frac{7}{15 \sqrt{2}}
\frac{1}{\sqrt{\pi}}$$
$$ \langle 6| \upsilon_s|6\rangle=\langle 8| \upsilon_s|8\rangle=\frac{2}{3}\sqrt{2}\frac{1}{\sqrt{\pi}},\langle 7| \upsilon_s|7\rangle=\sqrt{2} \frac{1}{\sqrt{\pi}}$$
$$ \langle 9| \upsilon_s|9\rangle=\langle 10| \upsilon_s|10\rangle=\frac{2}{3}\sqrt{2}\frac{1}{\sqrt{\pi}}$$
Combining these with the color factor given above for the  component $(\lambda,\mu)$ and the spin S of the 2-quark system we get:
\barr
\langle i,(\lambda',\mu')S'|V_{12} \upsilon_s|j(\lambda,\mu)S\rangle&=&\frac{\alpha_s}{b}C(\lambda,\mu)\delta_{\lambda,\lambda'}\delta_{\mu,\mu'}\delta_{S,S'}\langle i| \upsilon_s|j\rangle,\nonumber\\C(\lambda,\mu)&=&\left \{ \begin{array}{rc}-\frac{4}{3},(\lambda,\mu)=(0,1)\\\frac{2}{3},(\lambda,\mu)=(2,0)\\ \end{array}\right .
\earr
Using this information we can construct the following 2-body matrices:\\
$I=1$ states:\\
i) $J=0^+$ matrix. The order of the basis is:\\
$|1\rangle(20)0, |2\rangle(20)0,|3\rangle(20)0,|6\rangle(20)1,|9\rangle(01)0$\\
where the first number indicates the orbital, the second the color and last the spin S.
$$ \left(
\begin{array}{ccccc}
\frac{2 \sqrt{\frac{2}{\pi }}}{3}
& \frac{1}{3} \sqrt{\frac{2}{3
		\pi }} & \frac{1}{3}
\sqrt{\frac{2}{3 \pi }} & 0&0 \\
\frac{1}{3} \sqrt{\frac{2}{3 \pi
}} & \frac{11}{9 \sqrt{2 \pi }}
& \frac{1}{9 \sqrt{2 \pi }} & 0&0
\\
\frac{1}{3} \sqrt{\frac{2}{3 \pi
}} & \frac{1}{9 \sqrt{2 \pi }}
& \frac{11}{9 \sqrt{2 \pi }} &
0 &0\\
0 & 0 & 0 & -\frac{8
	\sqrt{\frac{2}{\pi }}}{9}&0 \\
0 & 0 & 0 &0 & \frac{4
	\sqrt{\frac{2}{\pi }}}{9} \\	
\end{array}
\right)$$
ii) $J=1^+$ matrix. The order of the basis is:\\
$|1\rangle(01)1, |2\rangle(01)1,|3\rangle(01)1,|4\rangle(01)1,|5\rangle(01)1,|6\rangle(01)1,|6\rangle(20)0,|9\rangle(20)1,|10\rangle(20)1$
$$\left(
\begin{array}{ccccccccc}
-\frac{4 \sqrt{\frac{2}{\pi
}}}{3} & -\frac{2}{3}
\sqrt{\frac{2}{3 \pi }} &
-\frac{2}{3} \sqrt{\frac{2}{3
		\pi }} & 0 & 0 & 0 & 0&0&0\\
-\frac{2}{3} \sqrt{\frac{2}{3 \pi
}} & -\frac{11
	\sqrt{\frac{2}{\pi }}}{9} &
-\frac{\sqrt{\frac{2}{\pi
}}}{9} & 0 & 0 & 0 & 0&0&0\\
-\frac{2}{3} \sqrt{\frac{2}{3 \pi
}} & -\frac{\sqrt{\frac{2}{\pi
}}}{9} & -\frac{11
	\sqrt{\frac{2}{\pi }}}{9} & 0 &
0 & 0 & 0 &0&0\\
0 & 0 & 0 & -\frac{46
	\sqrt{\frac{2}{\pi }}}{45} &
-\frac{14 \sqrt{\frac{2}{\pi
}}}{45} & 0 & 0&0&0 \\
0 & 0 & 0 & -\frac{14
	\sqrt{\frac{2}{\pi }}}{45} &
-\frac{46 \sqrt{\frac{2}{\pi
}}}{45} & 0 & 0&0&0 \\
0 & 0 & 0 & 0 & 0 & -\frac{8
	\sqrt{\frac{2}{\pi }}}{9} & 0&0&0\\
0 & 0 & 0 & 0 & 0 & 0 & \frac{4
	\sqrt{\frac{2}{\pi }}}{9}&0&0 \\
0 & 0 & 0 & 0 & 0 & 0 & 0&\frac{4	\sqrt{\frac{2}{\pi }}}{9}&0 \\
0 & 0 & 0 & 0 & 0 & 0 & 0&0&\frac{4	\sqrt{\frac{2}{\pi }}}{9}\\
\end{array}
\right) $$
ii) $J=2^+$ matrix. The order of the basis is:\\
$|4\rangle(01)1,|4\rangle(20)0,|5\rangle(01)1,|5\rangle(20)0,|6\rangle(20)1,|10\rangle(01)0,|10\rangle(20)1$
$$\left(
\begin{array}{ccccccc}
-\frac{46 \sqrt{\frac{2}{\pi
}}}{45} & 0 & -\frac{14
	\sqrt{\frac{2}{\pi }}}{45} & 0& 0&0&0\\
0 & \frac{23 \sqrt{\frac{2}{\pi
}}}{45} & 0 & \frac{7
	\sqrt{\frac{2}{\pi }}}{45} & 0&0&0\\
-\frac{14 \sqrt{\frac{2}{\pi
}}}{45} & 0 & -\frac{46
	\sqrt{\frac{2}{\pi }}}{45} & 0 & 0 &0&0\\
0 & \frac{7 \sqrt{\frac{2}{\pi
}}}{45} & 0 & \frac{23
	\sqrt{\frac{2}{\pi }}}{45} & 0&0&0\\
0 & 0 & 0 & 0 & \frac{4\sqrt{\frac{2}{\pi }}}{9}&0&0 \\
0 & 0 & 0 & 0 &0&- \frac{8\sqrt{\frac{2}{\pi }}}{9}&0 \\
0 & 0 & 0 & 0 &0&0&\frac{4\sqrt{\frac{2}{\pi }}}{9} \\
\end{array}
\right) $$
iv) $J=3^+$ matrix. The order of the basis is:\\
$|4\rangle(01)1,|5\rangle(01)1,|10\rangle(20)1,$
$$\left(
\begin{array}{ccc}
-\frac{46 \sqrt{\frac{2}{\pi
}}}{45} & -\frac{14	\sqrt{\frac{2}{\pi }}}{45}&0 \\
-\frac{14 \sqrt{\frac{2}{\pi
}}}{45} & -\frac{46	\sqrt{\frac{2}{\pi }}}{45}&0 \\
0&0&\frac{4\sqrt{\frac{2}{\pi }}}{9}\\
\end{array}
\right)$$
Matrices of negative parity states:
$$0^-=\left(
\begin{array}{cc}
-\frac{4 \sqrt{\frac{2}{\pi }}}{3} &
0 \\
0 & \frac{4 \sqrt{\frac{2}{\pi
}}}{9} \\
\end{array}
\right),\,1^-= \left(
\begin{array}{cccc}
-\frac{4 \sqrt{\frac{2}{\pi }}}{3} &
0 & 0 & 0 \\
0 & \frac{2 \sqrt{\frac{2}{\pi
}}}{3} & 0 & 0 \\
0 & 0 & -\frac{8 \sqrt{\frac{2}{\pi
}}}{9} & 0 \\
0 & 0 & 0 & \frac{4
	\sqrt{\frac{2}{\pi }}}{9} \\
\end{array}
\right),\,2^-=\left(
\begin{array}{cc}
-\frac{4 \sqrt{\frac{2}{\pi }}}{3} &
0 \\
0 & \frac{4 \sqrt{\frac{2}{\pi
}}}{9} \\
\end{array}
\right)$$
with the ordering of the bases being:
$$ |7\rangle(01)1,|8\rangle(20)1;\,|7\rangle(01)1,|7\rangle(20)0,|8\rangle(01)0,|8\rangle(20)1;\,|7\rangle(01)1,|8\rangle(20)1$$
respectively.

$I=0$ states:\\
i) $J=0^+$ matrix.  The order of the basis is:\\
$|1\rangle(01)0, |2\rangle(01)0,|3\rangle(01)0,|6\rangle(01)1,|9\rangle(01)1$\\
$$\left(
\begin{array}{ccccc}
-\frac{4 \sqrt{\frac{2}{\pi
}}}{3} & -\frac{2}{3}
\sqrt{\frac{2}{3 \pi }} &
-\frac{2}{3} \sqrt{\frac{2}{3\pi }} & 0&0 \\
-\frac{2}{3} \sqrt{\frac{2}{3 \pi
}} & -\frac{11	\sqrt{\frac{2}{\pi }}}{9} &
-\frac{\sqrt{\frac{2}{\pi	}}}{9} & 0&0 \\
-\frac{2}{3} \sqrt{\frac{2}{3 \pi
}} & -\frac{\sqrt{\frac{2}{\pi
}}}{9} & -\frac{11
	\sqrt{\frac{2}{\pi }}}{9} & 0&0\\
0 & 0 & 0 & -\frac{8
	\sqrt{\frac{2}{\pi }}}{9}&0 \\

0 & 0 & 0 &0& -\frac{8
	\sqrt{\frac{2}{\pi }}}{9}\\
\end{array}
\right)$$
ii) $J=1^+$. The order of the basis is:\\
$|1\rangle(20)1, |2\rangle(20)1,|3\rangle(20)1,|4\rangle(20)1,|5\rangle(20)1,|6\rangle(01)1,|6\rangle(20)0,|9\rangle(01)1,|10\rangle(01)1$
$$ \left(
\begin{array}{ccccccccc}
\frac{2 \sqrt{\frac{2}{\pi }}}{3}
& \frac{1}{3} \sqrt{\frac{2}{3
		\pi }} & \frac{1}{3}
\sqrt{\frac{2}{3 \pi }} & 0 & 0
& 0 & 0&0&0 \\
\frac{1}{3} \sqrt{\frac{2}{3 \pi
}} & \frac{11}{9 \sqrt{2 \pi }}
& \frac{1}{9 \sqrt{2 \pi }} & 0
& 0 & 0 &0&0&0 \\
\frac{1}{3} \sqrt{\frac{2}{3 \pi
}} & \frac{1}{9 \sqrt{2 \pi }}
& \frac{11}{9 \sqrt{2 \pi }} &
0 & 0 & 0 & 0 &0&0\\
0 & 0 & 0 & \frac{23
	\sqrt{\frac{2}{\pi }}}{45} &
\frac{7 \sqrt{\frac{2}{\pi
}}}{45} & 0 & 0&0&0 \\
0 & 0 & 0 & \frac{7
	\sqrt{\frac{2}{\pi }}}{45} &
\frac{23 \sqrt{\frac{2}{\pi
}}}{45} & 0 & 0&0&0 \\
0 & 0 & 0 & 0 & 0 & -\frac{8
	\sqrt{\frac{2}{\pi }}}{9} & 0
\\
0 & 0 & 0 & 0 & 0 & 0 & \frac{4
	\sqrt{\frac{2}{\pi }}}{9}&0&0 \\
0 & 0 & 0 & 0 & 0 & 0 &0& -\frac{8
	\sqrt{\frac{2}{\pi }}}{9}&0 \\
0 & 0 & 0 & 0 & 0 & 0 &0&0& -\frac{8
	\sqrt{\frac{2}{\pi }}}{9} \\
\end{array}
\right)$$
iii) $J=2^+$The order of the basis is:\\
$|4\rangle(01)0,|4\rangle(20)1,|5\rangle(01)0,|5\rangle(20)1,|6\rangle(01)1,|10\rangle(01)1,|10\rangle(20)0$\\
$$ \left(
\begin{array}{ccccccc}
-\frac{46\sqrt{\frac{2}{\pi
}}}{45} & 0 & -\frac{14
	\sqrt{\frac{2}{\pi }}}{45} & 0
& 0&0&0 \\
0 & \frac{23 \sqrt{\frac{2}{\pi
}}}{45} & 0 & \frac{7
	\sqrt{\frac{2}{\pi }}}{45} & 0&0&0
\\
-\frac{14 \sqrt{\frac{2}{\pi
}}}{45} & 0 & -\frac{46
	\sqrt{\frac{2}{\pi }}}{45} & 0
& 0 &0&0\\
0 & \frac{7 \sqrt{\frac{2}{\pi
}}}{45} & 0 & \frac{23
	\sqrt{\frac{2}{\pi }}}{45} & 0&0&0
\\
0 & 0 & 0 & 0 & -\frac{8
	\sqrt{\frac{2}{\pi }}}{9}&0&0\\
0 & 0 & 0 & 0 &0& -\frac{8
	\sqrt{\frac{2}{\pi }}}{9}&0\\
0 & 0 & 0 & 0 &0&0 &\frac{4
	\sqrt{\frac{2}{\pi }}}{9}\\
\end{array}
\right)$$
iv) $J=3^+$. The order of the basis is:\\
$|4\rangle(20)1,|5\rangle(20)1,|10\rangle(01)1$
$$\left(
\begin{array}{ccc}
\frac{23 \sqrt{\frac{2}{\pi
}}}{45} & \frac{7
	\sqrt{\frac{2}{\pi }}}{45} &0\\
\frac{7 \sqrt{\frac{2}{\pi
}}}{45} &\frac{23
	\sqrt{\frac{2}{\pi }}}{45}&0 \\
0&0&-\frac{8
	\sqrt{\frac{2}{\pi }}}{9}\\
\end{array}
\right) $$
Matrices of negative parity states:
$$0^-=\left(
\begin{array}{cc}
\frac{2 \sqrt{\frac{2}{\pi }}}{3} &
0 \\
0 & -\frac{8 \sqrt{\frac{2}{\pi
}}}{9} \\
\end{array}
\right),\,1^-=\left(
\begin{array}{cccc}
-\frac{4 \sqrt{\frac{2}{\pi }}}{3} &
0 & 0 & 0 \\
0 & \frac{2 \sqrt{\frac{2}{\pi
}}}{3} & 0 & 0 \\
0 & 0 & -\frac{8 \sqrt{\frac{2}{\pi
}}}{9} & 0 \\
0 & 0 & 0 & \frac{4
	\sqrt{\frac{2}{\pi }}}{9} \\
\end{array}
\right),\,2^-=\left(
\begin{array}{cc}
\frac{2 \sqrt{\frac{2}{\pi }}}{3} &
0 \\
0 & -\frac{8 \sqrt{\frac{2}{\pi
}}}{9} \\
\end{array}
\right)$$
with the ordering of the bases being:
$$ |7\rangle(20)1,|8\rangle(01)1;\,|7\rangle(01)0,|7\rangle(20)1,|8\rangle(01)1,|8\rangle(20)0;\,|7\rangle(20)1,|8\rangle(01)1$$
respectively.

The above matrices must, of course, be multiplied by the energy scale factor sc=$\alpha_s\frac{\hbar c }{b_q}$. The nucleon radius $R_=\sqrt{\langle r^2\rangle}$ is estimated to be 0.8 F.  This leads to $b_q=0.65$ F and sc$\approx$ 300 MeV.
\clearpage
\newpage

\section{Appendix B: some  mathematical tools for the structure of the six-quarks}
\label{sec.Appendix B}
Those involved in the symmetry $f_{cs}$ to $SU_c(3)\times SU_I(2)$ can be obtained from the one particle CFP's 
calculated by  So and Strottman \cite{SOSTRO}.

\begin{table}[h!]
	\begin{center}
		\caption{ The one particle CFP's involving the color-spin symmetry, needed in the case of a six-quark cluster,  color singlet with  spins 0,1,2  and 3.}
		\label{tab:spc6cfp5}
		$$
		\begin{array}{|c|c|c|}
		\hline
		[f_1](\lambda_1,\mu_1)S_1&[f](\lambda,\mu)S&\mbox{CFP}\\
		\hline
		\mbox{$[2^2,1](0,1)\frac{1}{2}$}&[2^3](00)1&\frac{\sqrt{5}}{3}\\
		\mbox{$[2^2,1](0,1)\frac{3}{2}$}&[2^3](00)1&\frac{2}{3}\\
		\hline
		\end{array}\quad
		\begin{array}{|c|c|c|}
		\hline
		[f_1](\lambda_1,\mu_1)S_1&[f](\lambda,\mu)S&\mbox{CFP}\\
		\hline
		\mbox{$[2^2,1](0,1)(0,1)\frac{5}{2}$}&[2^3](00)3&1\\
		\hline
		\end{array}
		$$
		$$
		\begin{array}{|c|c|c|}
		\hline
		[f_1](\lambda_1,\mu_1)S_1&[f](\lambda,\mu)S&\mbox{CFP}\\
		\hline
		\mbox{$[2^2,1](0,1)\frac{1}{2}$}&[2^21^2](00)0&1\\
		\hline
		\end{array}\quad
		\begin{array}{|c|c|r|}
		\hline
		[f_1](\lambda_1,\mu_1)S_1&[f](\lambda,\mu)S&\mbox{CFP}\\
		\hline
		\mbox{$[2^2,1](0,1)\frac{3}{2}$}&[2^21^2](00)2&\frac{4}{5}\\
		\mbox{$[2^2,1](0,1)\frac{5}{2}$}&[2^2,1^2](00)2&-\frac{3}{5}\\
		\hline
		\end{array}
		$$
	\end{center}
\end{table} 
\begin{table}[h!]
	\begin{center}
		\caption{ The two particle CFP's involving the color-spin symmetry, needed in the case of a 6-quark cluster,  color singlet with  spin zero and 2.}
		\label{tab:spc6cfp3}
		$$
		\begin{array}{|c|c|c|c|}
		\hline
		[f_1](\lambda_1,\mu_1)S_1&[f_2](\lambda_2,\mu_2)S_2&[f](\lambda,\mu)S&\mbox{CFP}\\
		\hline
		\mbox{$[2,2](0,2)0$}&[1^2](2,0)0&[2^21^2](00)0&-\frac{\sqrt{3}}{2}\\
		\mbox{$[2,2](1,0)1$}&[1^2](0,1)1&[2^21^2](00)0&\frac{1}{2}\\
		\hline
		\end{array}
		\begin{array}{|c|c|c|c|}
		\hline
		[f_1](\lambda_1,\mu_1)S_1&[f_2](\lambda_2,\mu_2)S_2&[f](\lambda,\mu)S&\mbox{CFP}\\
		\hline
		\mbox{$[1^4](0,2)0$}&[1^2](2,0)0&[2^21^2](00)0&-\sqrt{\frac{3}{5}}\\
		\mbox{$[1^4](1,0)1$}&[1^2](0,1)1&[2^21^2](00)0&-\sqrt{\frac{2}{5}}\\
		\hline
		\end{array}
		$$
		$$
		\begin{array}{|c|c|c|c|}
		\hline
		[f_1](\lambda_1,\mu_1)S_1&[f_2](\lambda_2,\mu_2)S_2&[f](\lambda,\mu)S&\mbox{CFP}\\
		\hline
		\mbox{$[2,1^2](0,2)1$}&[2](2,0)1&[2^21^2](00)0&\frac{1}{\sqrt{2}}\\
		\mbox{$[2,1^2](0,2)1$}&[2](2,0)1&[2^21^2](00)0&-\frac{1}{\sqrt{2}}\\
		\hline
		\end{array}
		\begin{array}{|c|c|c|c|}
		\hline
		[f_1](\lambda_1,\mu_1)S_1&[f_2](\lambda_2,\mu_2)S_2&[f](\lambda,\mu)S&\mbox{CFP}\\
		\hline
		%\mbox{$[2,1^2](0,2)1$}&[1^2](2,0)0&[2^21^2](00)0&0\\
		\mbox{$[2,1^2](0,2)1$}&[1^2](0,1)1&[2^21^2](00)2&1\\
		\hline
		\end{array}
		$$
	\end{center}
\end{table}
%%%%%%%%%%%%%%%%%%%%%%%%%%%%%%%%%%%
\begin{table}[h!]
	\caption{the  $[f_1](\lambda_1\mu_1)S_1,[f_2](\lambda_2\mu_2)S_2| [2^21^2](0,0)0$ full cfp}
	\label{tab:cfp2211}
$$
\left(
\begin{array}{ccccccc}
\mbox{$[f_1]$}&(\lambda_1\mu_1)&S_1,&\mbox{$[f_2]$}&(\lambda_2\mu_2)&S_2&\mbox{cfp}\\
\mbox{$[2^2]$} & (0,2) & 0, &\mbox{$[1^2]$}  &(2,0) & 0 & -\frac{1}{2} \\
\mbox{$[2^2]$} & (1,0) & 1 ,&\mbox{$[1^2]$} &(0,1) & 1 & \frac{1}{2\sqrt{3}} \\
\mbox{$[1^4]$} & (0,2) & 0 ,&\mbox{$[1^2]$} & (2,0) & 0 &\frac{1}{\sqrt{10}} \\
\mbox{$[1^4]$} & (1,0) & 1 ,&\mbox{$[1^2]$} & (0,1) & 1 &\frac{1}{\sqrt{15}} \\
\mbox{$[2,1^2]$} & (0,2) & 1, &\mbox{$[2]$} & (2,0) & 1 &\frac{1}{2} \\
\mbox{$[2,1^2]$} & (1,0) & 0. &\mbox{$[2]$} & (0,1) & 0 &-\frac{1}{2} \\
\end{array}
\right)
$$
\end{table}
% New Table%%%%%%%%%%%%%%%%%%%%%%%%%%%%%%%%%%%%%%%%%%%%%%%%%%%%%%%%%%
\begin{table}[h!]
	\begin{center}
		%\caption{ The two particle sc 2  cfp's needed in the case of the 6-quark cluster for $S=2$}
		\caption{ The two particle CFP's involving the color-spin symmetry, needed in the case of a 6-quark cluster,  color singlet with  spin two.}
		\label{tab:spc6cfp4}
		$$
		\begin{array}{|c|c|c|c|}
		\hline
		[f_1](\lambda_1,\mu_1)S_1&[f_2](\lambda_2,\mu_2)S_2&[f](\lambda,\mu)S&\mbox{CFP}\\
		\hline
		\mbox{$[2,2](0,2)2$}&[1^2](2,0)0&[2^21^2](00)2&\sqrt{\frac{3}{5}}\\
		\mbox{$[2,2](1,0)1$}&[1^2](0,1)1&[2^21^2](00)2&-\sqrt{\frac{2}{5}}\\
		\hline
		\end{array}
		\begin{array}{|c|c|c|c|}
		\hline
		[f_1](\lambda_1,\mu_1)S_1&[f_2](\lambda_2,\mu_2)S_2&[f](\lambda,\mu)S&\mbox{CFP}\\
		\hline
		\mbox{$[2,1^2](0,2)1$}&[2](2,0)1&[2^21^2](00)2&-\sqrt{\frac{4}{5}}\\
		\mbox{$[2,1^2](1,0)2$}&[2](0,1)0&[2^21^2](00)0&-\sqrt{\frac{1}{5}}\\
		\hline
		\end{array}
		$$
		$$
		\begin{array}{|c|c|c|c|}
		\hline
		[f_1](\lambda_1,\mu_1)S_1&[f_2](\lambda_2,\mu_2)S_2&[f](\lambda,\mu)S&\mbox{CFP}\\
		\hline
		\mbox{$[2,1^2](1,0)1$}&[1^2](0,1)1&[2^21^2](00)2&\sqrt{\frac{2}{5}}\\
		\mbox{$[2,1^2](1,0)2$}&[1^2](0,1)2&[2^21^2](00)2&-\sqrt{\frac{3}{5}}\\
		\hline
		\end{array}
		\begin{array}{|c|c|c|c|}
		\hline
		[f_1](\lambda_1,\mu_1)S_1&[f_2](\lambda_2,\mu_2)S_2&[f](\lambda,\mu)S&\mbox{CFP}\\
		\hline
		\mbox{$[1^4](1,0)1$}&[1^2](0,1)1&[2^21^2](00)2&1\\
		%\mbox{$[2,1^2](0,2)1$}&[1^2](0,1)1&[2^21^2](00)01&1\\
		\hline
		\end{array}
		$$
	\end{center}
\end{table}

% New Table%%%%%%%%%%%%%%%%%%%%%%%%%%%%%%%%%%%%%%%%%%%%%%%%%%%%%%%%%%

\begin{table}[h!]
	\begin{center}
		\caption{ The two particle CFP's involving the color-spin symmetry, needed in the case of a 6-quark cluster,  color singlet with  spins 1 and 3.}
		\label{tab:spc6cfp5}
		$$
		\begin{array}{|c|c|c|c|}
		\hline
		[f_1](\lambda_1,\mu_1)S_1&[f_2](\lambda_2,\mu_2)S_2&[f](\lambda,\mu)S&\mbox{CFP}\\
		\hline
		\mbox{$[2,2](0,2)0$}&[2](2,0)1&[2^3](00)1&\sqrt{\frac{5}{12}}\\
		\mbox{$[2,2](0,2)2$}&[2](2,0)1&[2^3](00)1&\frac{1}{\sqrt{6}}\\
		\mbox{$[2,2](1,0)0$}&[2](0,1)0&[2^3](00)1&\sqrt{\frac{5}{12}}\\
		\hline
		\end{array}
		\begin{array}{|c|c|c|c|}
		\hline
		[f_1](\lambda_1,\mu_1)S_1&[f_2](\lambda_2,\mu_2)S_2&[f](\lambda,\mu)S&\mbox{CFP}\\
		\hline
		\mbox{$[2,1^2](0,2)1$}&[1^2](2,0)0&[2^3](00)1&-\sqrt{\frac{5}{18}}\\
		\mbox{$[2,1^2](1,0)0$}&[1^2](0,1)1&[2^3](00)1&-\sqrt{\frac{5}{54}}\\
		\mbox{$[2,1^2](1,0)1$}&[1^2](0,1)1&[2^3](00)1&-\sqrt{\frac{5}{9}}\\
		\mbox{$[2,1^2](1,0)2$}&[1^2](0,1)1&[2^3](00)1&-\sqrt{\frac{2}{27}}\\
		\hline
		\hline
		\end{array}
		$$
		$$
		\begin{array}{|c|c|c|c|}
		\hline
		[f_1](\lambda_1,\mu_1)S_1&[f_2](\lambda_2,\mu_2)S_2&[f](\lambda,\mu)S&\mbox{CFP}\\
		\hline
		\mbox{$[2^2](0,2)2$}&[2](2,0)1&[2^3](00)3&1\\
		%\mbox{$[2,1^2](1,0)2$}&[1^2](0,1)1&[2^21^2](00)3&-1\\
		\hline
		\end{array}
		\begin{array}{|c|c|c|c|}
		\hline
		[f_1](\lambda_1,\mu_1)S_1&[f_2](\lambda_2,\mu_2)S_2&[f](\lambda,\mu)S&\mbox{CFP}\\
		\hline
		%\mbox{$[2,1^2](0,2)1$}&[1^2](2,0)0&[2^21^2](00)0&0\\
		\mbox{$[2,1^2](1,0)2$}&[1^2](0,1)1&[2^3](00)3&1\\
		\hline
		\end{array}
		$$
	\end{center}
\end{table}
\clearpage
%From those we obtained the needed 2-particle CFP's which are given in tables \ref{tab:spc6cfp3} and \ref{tab:spc6cfp4}
\section{Appendix C: Radial overlaps}
\label{AppendixRaD}
$$\left(
\begin{array}{ccc}
\{n_1,&n_2\}&C^{0s0s}_{n_1,n_2}\\
%	&\begin{array}{ccc}
0 & 0 & 0.786527 \\
0 & 1 & 0.740996 \\
0 & 2 & 0.318638 \\
1 & 1 & 0.174525 \\
0 & 3 & 0.132372 \\
1 & 2 & 0.150096 \\
0 & 4 & 0.0540008 \\
1 & 3 & 0.0623548 \\
2 & 2 & 0.0322717 \\
0 & 5 & 0.0217833 \\
1 & 4 & 0.0254374 \\
2 & 3 & 0.0268134 \\
0 & 6 & 0.0087203 \\
6 & 0 & 0.0087203 \\
1 & 5 & 0.0102611 \\
2 & 4 & 0.0109384 \\
3 & 3 & 0.00556957 \\
0 & 7 & 0.00347168 \\
1 & 6 & 0.00410775 \\
2 & 5 & 0.00441243 \\
3 & 4 & 0.00454417 \\
0 & 8 & 0.00137636 \\
1 & 7 & 0.00163535 \\
2 & 6 & 0.00176638 \\
3 & 5 & 0.00183306 \\
4 & 4 & 0.000926888 \\
0 & 9 & 0.000543873 \\
1 & 8 & 0.00064834 \\
2 & 7 & 0.000703224 \\
3 & 6 & 0.000733813 \\
\end{array}
\right),\,
\left(
\begin{array}{ccc}
\{n_1,&n_2\}&C^{0s0s}_{n_1,n_2}\\
4 & 5 & 0.000747791 \\
0 & 10 & 0.000214348 \\
1 & 9 & 0.000256195 \\
2 & 8 & 0.000278795 \\
3 & 7 & 0.000292142 \\
4 & 6 & 0.000299356 \\
5 & 5 & 0.000150825 \\
0 & 11 & 0.0000842943 \\
1 & 10 & 0.00010097 \\
2 & 9 & 0.000110167 \\
3 & 8 & 0.00011582 \\
4 & 7 & 0.000119178 \\
6 & 5 & 0.000120757 \\
0 & 12 & 0.0000330894 \\
1 & 11 & 0.0000397073 \\
2 & 10 & 0.0000434183 \\
3 & 9 & 0.0000457669 \\
4 & 8 & 0.0000472484 \\
5 & 7 & 0.000048075 \\
6 & 6 & 0.0000241707 \\
0 & 13 & 0.0000129691 \\
1 & 12 & 0.000015587 \\
2 & 11 & 0.0000170747 \\
3 & 10 & 0.0000180374 \\
4 & 9 & 0.0000186704 \\
5 & 8 & 0.0000190595 \\
6 & 7 & 0.0000192454 \\
\end{array}
\right)
$$	
$$
\left(
\begin{array}{ccc}
\{n_1,&n_2\}&C^{0s1s}_{n_1,n_2}\\
0 & 0 & 0.670177 \\
0 & 1 & 0.482225 \\
1 & 0 & 0.315691 \\
0 & 2 & 0.278207 \\
2 & 0 & 0.135751 \\
1 & 1 & 0.227155 \\
0 & 3 & 0.144882 \\
3 & 0 & 0.0563954 \\
1 & 2 & 0.131051 \\
2 & 1 & 0.0976796 \\
0 & 4 & 0.0710344 \\
4 & 0 & 0.0230063 \\
1 & 3 & 0.0682477 \\
3 & 1 & 0.0405793 \\
2 & 2 & 0.0563536 \\
0 & 5 & 0.0334612 \\
5 & 0 & 0.00928046 \\
4 & 1 & 0.0165542 \\
2 & 3 & 0.0293474 \\
3 & 2 & 0.0234111 \\
0 & 6 & 0.015318 \\
6 & 0 & 0.00371516 \\
1 & 5 & 0.0157621 \\
5 & 1 & 0.00667775 \\
2 & 4 & 0.0143887 \\
4 & 2 & 0.00955047 \\
3 & 3 & 0.0121918 \\
0 & 7 & 0.00686343 \\
7 & 0 & 0.00147906 \\
1 & 6 & 0.00721562 \\
6 & 1 & 0.00267324 \\
2 & 5 & 0.0067779 \\
5 & 2 & 0.00385255 \\
3 & 4 & 0.00597754 \\
4 & 3 & 0.00497362 \\
0 & 8 & 0.00302425 \\
8 & 0 & 0.000586376 \\
1 & 7 & 0.00323306 \\
7 & 1 & 0.00106426 \\
2 & 6 & 0.00310281 \\
6 & 2 & 0.00154225 \\
3 & 5 & 0.00281576 \\
5 & 3 & 0.0020063 \\
4 & 4 & 0.00243852 \\
\end{array}
\right)
,\,
\left(
\begin{array}{ccc}
\{n_1,&n_2\}&C^{0s0d}_{n_1,n_2}\\
%\left(
%\begin{array}{ccc}
0 & 0 & 0.670177 \\
0 & 1 & 0.482225 \\
1 & 0 & 0.315691 \\
0 & 2 & 0.278207 \\
2 & 0 & 0.135751 \\
1 & 1 & 0.227155 \\
0 & 3 & 0.144882 \\
3 & 0 & 0.0563954 \\
1 & 2 & 0.131051 \\
2 & 1 & 0.0976796 \\
0 & 4 & 0.0710344 \\
4 & 0 & 0.0230063 \\
1 & 3 & 0.0682477 \\
3 & 1 & 0.0405793 \\
2 & 2 & 0.0563536 \\
0 & 5 & 0.0334612 \\
5 & 0 & 0.00928046 \\
4 & 1 & 0.0165542 \\
2 & 3 & 0.0293474 \\
3 & 2 & 0.0234111 \\
0 & 6 & 0.015318 \\
6 & 0 & 0.00371516 \\
1 & 5 & 0.0157621 \\
5 & 1 & 0.00667775 \\
2 & 4 & 0.0143887 \\
4 & 2 & 0.00955047 \\
3 & 3 & 0.0121918 \\
0 & 7 & 0.00686343 \\
7 & 0 & 0.00147906 \\
1 & 6 & 0.00721562 \\
6 & 1 & 0.00267324 \\
2 & 5 & 0.0067779 \\
5 & 2 & 0.00385255 \\
3 & 4 & 0.00597754 \\
4 & 3 & 0.00497362 \\
0 & 8 & 0.00302425 \\
8 & 0 & 0.000586376 \\
1 & 7 & 0.00323306 \\
7 & 1 & 0.00106426 \\
2 & 6 & 0.00310281 \\
6 & 2 & 0.00154225 \\
3 & 5 & 0.00281576 \\
5 & 3 & 0.0020063 \\
4 & 4 & 0.00243852 \\
\end{array}
\right)
$$
\section{Appendix D: Formulae for the expansion of six-quark to $3q\times 3q$
\label{AppendixD}}

i). We begin with $0s^6$ configuration.

What we want to do is perform the expansions:
\barr
|0s^6[2^2,1^2](0,0)0,1\rangle&=&\sum \langle [f_1](\lambda,\mu)]S_1, I_1,[f_1](\mu,\lambda)]S_1 ,I_2|[f](0,0)S,I\rangle\nonumber\\
&&| [f_1](\lambda,\mu)]S_1, I_1\rangle| [f_1](\mu,\lambda)]S_1 ,I_2\rangle
\earr
\barr
|0s^6[2^3](0,0)1,0\rangle&=&\sum \langle [f_1](\lambda,\mu)]S_1, I_1,[f_1](\mu,\lambda)]S_2 ,I_1|[2^3](0,0)0,1\rangle\nonumber\\
&& |[f_1](\lambda,\mu)]S_1, I_1\rangle| [f_1](\mu,\lambda)]S_2 ,I_1\rangle
\earr
In the present case due to isospin restrictions the allowed possibilities are $f_1=[2,1]$ and $f_1=[1^3]$. We are interested in nucleon like components, i.e. only
$f_1=[2,1](0,0)\frac{1}{2},\frac{1}{2}$,$f_1=[2,1](1,1)\frac{1}{2},\frac{1}{2}$ and $f_1=[2,1](1,1)\frac{3}{2},\frac{1}{2}$

the needed CFPs can be calculated the usual way:
\barr
&&\langle [2,1](\lambda,\mu)]S_1, I_1,[2,1](\mu,\lambda)]S_2 ,I_2|[f](0,0)S,I\rangle U([21][1][f][f'][f_1][f_2])\nonumber\\
&=&\sum \langle [21](\lambda,\mu)]S_1, I_1,[1](1,0)\frac{1}{2}\frac{1}{2}|[f']((\mu',\lambda')S',I'\rangle\nonumber\\
& &\langle [f'](\mu',\lambda')]S', I',[f_2](\lambda',\mu')S'_2,I'_2[f]((0,0),S,I\rangle\nonumber\\
&&\langle[1](1,0))\frac{1}{2}\frac{1}{2},[1](1,0))\frac{1}{2}\frac{1}{2}|[f_2](\lambda',\mu')S_2,I_2\rangle\nonumber\\
&& U(S_1,\frac{1}{2},S,S'_2,S',S_2)U(I_1,\frac{1}{2},I,I'_2,I',I_2)U((\lambda,\mu),(1,0)(00),(\lambda',\mu'),(\lambda'.\mu')(\mu,\lambda))\nonumber\\
\earr
In the above expression [f'] can be selected arbitrarily among the allowed values. We have chosen the symmetry $[2^2]$. We consider the two cases of interest to us, namely   $f=[2^2,1^2]$ for $I=1$ and  $[f]=[2^3]$ for $I=0$. With these restrictions the summation extends over all indices, which do not appear on the left hand side of the equation.

We begin with  $J=0,\,I=1$ 6q cluster. We need the following two particle cfp's:
$$ \langle [2^2])(0,2)0,[1^2](2,0)0||[2^21^2](00)0\rangle=\frac{\sqrt{3}}{2},\,\langle [2^2])(1,0)1,[1^2](0,1)1||[2^21^2](0,0)0\rangle=-\frac{1}{2}$$
Then we get the 3-particle cfp's
$$\langle[2,1](00)1/21/2,[2,1](00)1/21/2|2^21^2(0,0)0,1\rangle =\frac{1}{2},$$ $$\langle[2,1](1,1)1/21/2,[2,1](1,1)1/21/2|2^21^2(0,0)0,1\rangle =\frac{1}{\sqrt{2}},$$ $$\langle[2,1](1,1)3/21/2,[2,1](1,1)3/21/2|2^21^2(0,0)0,1\rangle =-\frac{1}{2}$$
Writing:	
\beq
{0s^6}_{IS}=\kappa^{0s^6}_{IS} \left [ [2,1](00)\frac{1}{2}\times [2,1](00)\frac{1}{2}\right ]IS 
\eeq
we find
\beq 
\kappa^{0s^6}_{1,0}=\frac{1}{2}
\eeq
On the other hand for the $I=0,\,J=1$ we use the following two particle cfp:
$$\langle [2^2])(0,2)0,[2](2,0)1||[2^3](00)1\rangle=\frac{\sqrt{5}}{2\sqrt{3}},\,\langle [2^2])(02)2,[2](2,0)1||[2^3](0,0)1\rangle=\frac{1}{\sqrt{6}}$$
$$\langle [2^2])(1,0)1,[2](0,1)0||[2^3](00)1\rangle=\frac{\sqrt{5}}{2\sqrt{3}} $$
Employing these we get the following 3-particle cfp's
$$\langle[2,1](00)1/21/2,[2,1](00)1/21/2|[2^3](0,0)1,0\rangle =-\frac{\sqrt{5}}{6},$$ $$\langle[2,1](1,1)1/21/2,[2,1](1,1)1/21/2|[2^3](0,0)1,0\rangle =-\frac{\sqrt{5}}{3\sqrt{2}},$$ $$\langle[2,1](1,1)3/21/2,[2,1](1,1)3/21/2|[2^3](0,0)1,0\rangle =\frac{1}{6},$$ $$\langle[2,1](1,1)3/21/2,[2,1](1,1)1/21/2|[2^3](0,0)1,0\rangle =\frac{\sqrt{5}}{3\sqrt{2}}, $$
$$ \langle[2,1](1,1)1/21/2,[2,1](1,1)3/21/2|[2^3](0,0)1,0\rangle =\frac{\sqrt{5}}{3\sqrt{2}}$$
In this case
\beq 
\kappa^{0s^6}_{0,1}=\frac{\sqrt{5}}{6}
\eeq
%	We see that in both cases  the two nucleon like component is fairly large.

%	$$ 0s^6=C{f}\left [ [2,1](00)\frac{1}{2}\times [2,1](00)\frac{1}{2}\right ]I,S,$$ $$ \mbox{ with } (I,S)=(1,0)\mbox{ for}[f]=[2^21^2] \mbox{ and} (I,S)=(0,1)\mbox{ for }[f]=[2^3]$$

ii) $0s^51s(0d)$ Now we have to perform the re-coupling
$$\left [0s^5 ([f_1]S_1,I_1(01))[1](10)\frac{1}{2}(10)\right ] (I,S)=\kappa^{0s^51s(0d)}_ {([f_1]S_1,I_1(01))}\left [ [2,1](00)\frac{1}{2}\times [2,1](00)\frac{1}{2}\right ]I,S$$
The needed coefficients can be calculated using the 2-particle coefficients $\langle 3,2|5\rangle $, found in the tables quark-new-to-use and then by angular momentum re-coupling, both in spin and isospin space. Note that only states with $S=0,1$ can be considered.
The results are shown  in tables 	\ref{tab:taba} and \ref{tab:tabb}, writing $\kappa^{0s^51s(0d)}_ { NN}$ for simplicity.\\
iii) $0s^41 0p^2$ Now we have to perform the re-coupling
$$\left [0s^4 ([f_1]S_1,(\mu,\lambda))I_1,[f_2](\lambda.\mu)S_2,I_2\right ] (I,S)=$$ $$\kappa^{4}_{0s^4 ([f_1]S_1,(\mu,\lambda))I_1[f_2](\lambda.\mu)S_2,I_2}\left [ [2,1](00)\frac{1}{2}\times [2,1](00)\frac{1}{2}\right ]I,S$$
The needed coefficients can be calculated using the $\langle 3,1|4\rangle $one particle cfp's, tabulated by So and Strottman \cite{SOSTRO}, and then by angular momentum re-coupling, both in spin and isospin space. Note that only the $2p^2$ states with color spun symmetry $(01)0$ and $(01)1$ can lead to 2-nucleon like configurations.

%\tableofcontents
\end{document}